\documentclass[twocolumn,english,showpacs,preprintnumbers,amsmath,amssymb,floatfix,nofootinbib]{revtex4-1}

\usepackage[normalem]{ulem}

\usepackage{tikz,xcolor}
\usepackage[colorlinks = true,
linkcolor = blue,
urlcolor  = blue,
citecolor = blue,
anchorcolor = blue]{hyperref}

\definecolor{lime}{HTML}{A6CE39}
\DeclareRobustCommand{\orcidicon}{%
	\begin{tikzpicture}
		\draw[lime, fill=lime] (0,0)
		circle [radius=0.16]
		node[white] {{\fontfamily{qag}\selectfont \tiny ID}};
		\draw[white, fill=white] (-0.0625,0.095)
		circle [radius=0.007];
	\end{tikzpicture}
	\hspace{-2mm}
}
\foreach \x in {A, ..., Z}{%
	\expandafter\xdef\csname orcid\x\endcsname{\noexpand\href{https://orcid.org/\csname orcidauthor\x\endcsname}{\noexpand\orcidicon}}
}

\usepackage[T1]{fontenc}
\usepackage[latin9]{inputenc}
\usepackage{color}
\usepackage{array}
\usepackage{amstext}
\usepackage{subfig}
\usepackage{graphicx}
\usepackage{esint}
\usepackage{rotating}
\usepackage{slashed}
\usepackage{siunitx}
\usepackage[normalem]{ulem}
\usepackage[font=small,labelfont=bf]{caption}
\usepackage{lipsum}
\usepackage{xcolor}

\newcommand{\xpom}{x_{\xpom}}

\makeatletter

\@ifundefined{textcolor}{}
{%
 \definecolor{BLACK}{gray}{0}
 \definecolor{WHITE}{gray}{1}
 \definecolor{RED}{rgb}{1,0,0}
 \definecolor{GREEN}{rgb}{0,1,0}
 \definecolor{BLUE}{rgb}{0,0,1}
 \definecolor{CYAN}{cmyk}{1,0,0,0}
\definecolor{MAGENTA}{cmyk}{0,1,0,0}
 \definecolor{YELLOW}{cmyk}{0,0,1,0}
 }

\@ifundefined{definecolor}
{\usepackage{color}}{}
\@ifundefined{definecolor}
{\usepackage{color}}{}
\makeatother
\usepackage{babel}

\def\Re{{\cal R \mskip-4mu \lower.1ex \hbox{\it e}\,}}
\def\Im{{\cal I \mskip-5mu \lower.1ex \hbox{\it m}\,}}

\def\tev{\,{\ifmmode\mathrm {TeV}\else TeV\fi}}
\def\gev{\,{\ifmmode\mathrm {GeV}\else GeV\fi}}
\def\mev{\,{\ifmmode\mathrm {MeV}\else MeV\fi}}

\begin{document}

%
%
\title{Determination of $K^0_S$ Fragmentation Functions including {\tt BESIII} Measurements and using Neural Networks  }
%
%

\author{Maryam~Soleymaninia$^{1}$\orcidB{}}
\email{Maryam\_Soleymaninia@ipm.ir}

\author{Hadi~Hashamipour$^{2,1}$\orcidA{}}
\email{ hadi.hashamipour@lnf.infn.it}

\author{Maral~Salajegheh$^{3}$\orcidC{}} 
\email{Maral@hiskp.uni-bonn.de }

\author{Hamzeh~Khanpour$^{4,5,1}$\orcidE{}}
\email{Hamzeh.Khanpour@cern.ch}

\author{Hubert Spiesberger$^6$}
\email{spiesber@uni-mainz.de}

\author{Ulf-G.~Mei{\ss}ner$^{3,7,8}$\orcidD{}}
\email{meissner@hiskp.uni-bonn.de }

\affiliation {
$^{1}$School of Particles and Accelerators, Institute for Research in Fundamental Sciences (IPM), P.O.Box 19395-5531, Tehran, Iran.   \\
$^{2}$Istituto Nazionale di Fisica Nucleare, Gruppo collegato di Cosenza,
I-87036 Arcavacata di Rende, Cosenza, Italy.   \\
$^{3}$Helmholtz-Institut f\"ur Strahlen-und Kernphysik and Bethe Center for Theoretical Physics, Universit\"at Bonn, D-53115 Bonn, Germany.  \\
$^{4}$AGH University, Faculty of Physics and Applied Computer Science, Al. Mickiewicza 30, 30-055 Kraków, Poland. \\ 
$^{5}$Department of Physics, University of Science and Technology of Mazandaran, P.O.Box 48518-78195, Behshahr, Iran.  \\
$^{6}$PRISMA{\color{red}$^{+}$} Cluster of Excellence, Institut f\"ur Physik, Johannes-Gutenberg-Universit\"at, Staudinger Weg 7, D-55099 Mainz, Germany \\
$^{7}$Institute for Advanced Simulation (IAS-4), Forschungszentrum J\"ulich, D-52425 J\"ulich, Germany. \\
$^{8}$Tbilisi State University, 0186 Tbilisi, Georgia.
}

\date{\today}

%
\begin{abstract}

In this study, we revisit the extraction of parton-to-$K^0_S$ hadron 
fragmentation functions, named \texttt{FF24-}$K^0_S$, focusing on 
both next-to-leading-order and next-to-next-to-leading-order accuracy 
in perturbative QCD. Our approach involves the analysis of single 
inclusive electron-positron annihilation (SIA) data. The two key 
improvements are, on the one hand, the incorporation of the latest 
experimental data from the {\tt BESIII} experiment and, on the other hand, the 
adoption of Neural Networks in the fitting procedure. To address 
experimental uncertainties, the Monte Carlo method is employed.
Our investigation also explores the impact of hadron mass corrections 
on the description of SIA data, spanning a broad kinematic regime with 
a particular emphasis on the range of small $z$ values. The theory 
prediction for $K^0_S$ production at both NLO and NNLO accuracy 
exhibits good agreement with experimental data within their respective 
uncertainties. 

\end{abstract}
%

\maketitle
\tableofcontents{}

%
\section{Introduction}
\label{sec:introduction}
%

Fragmentation functions (FFs) play a crucial role in the computation of 
scattering cross sections that involve detected final-state hadrons. 
When there is a high energy scale, QCD factorization allows us to 
disentangle the physics of a parton transitioning to a colorless hadron 
from the hard scattering process that generated the parton~\cite{Collins}. 
Parton-to-hadron FFs precisely depict this transition and illustrate how 
QCD final states are nonperturbatively created through hadronization, 
providing insight into the dynamics of the strong interaction. 
Fragmentation functions are process-independent quantities and can be 
determined from data through a comprehensive QCD analysis. The  variation of
the FFs with energy can be perturbatively calculated, comprising an 
expansion in the strong coupling, and is presently understood up to 
next-to-next-to-leading order (NNLO) 
accuracy~\cite{Almasy:2011eq,Mitov:2006ic}. Due to the significance of 
FFs, there has been considerable theoretical attention towards their 
global analysis, resulting in the development of multiple new sets 
for various hadron FFs~\cite{Soleymaninia:2022alt,
AbdulKhalek:2022laj,Borsa:2022vvp,Moffat:2021dji}.

Fragmentation functions have a large diversity of applications. 
Their knowledge is essential in numerous hard scattering processes 
at both existing and upcoming experiments, such as at the 
Large Hadron Collider (LHC), the Future Circular Collider 
(FCC)~\cite{FCC:2018byv,FCC:2018bvk}, and the Electron-Ion Collider 
(EIC)~\cite{AbdulKhalek:2021gbh}. Particle production measurements 
provide valuable insight into the dynamics of QCD interactions at 
low momentum transfer. Accurate modeling of these interactions is 
crucial for understanding and constraining the impact of the underlying 
event in collisions at high transverse momentum investigated at the 
LHC. Other applications include the examination of the spin structure 
of the nucleon~\cite{Borsa:2020lsz}. Moreover, FFs play a pivotal role 
in investigating hadron production rates in scattering processess 
involving heavy nuclei~\cite{Klasen:2023uqj}.

The distribution of hadrons containing strange quarks has been 
documented at various center-of-mass energies, both at the 
LHC~\cite{ATLAS:2019pqg,ATLAS:2011xhu} and at the 
Tevatron~\cite{CDF:2005ofj}. Kaons are particularly important. 
FFs for $K^0_S$ production have been calculated by some of the 
authors of the present paper~\cite{Soleymaninia:2020ahn}, denoted 
\texttt{SAK20}. Other research groups, such as 
\texttt{BKK96}~\cite{Binnewies:1995kg}, 
\texttt{BS}~\cite{Bourrely:2003wi}, 
\texttt{AKK05}~\cite{Albino:2005mv}, 
\texttt{AKK08}~\cite{Albino:2008fy},  
\texttt{DSS17}~\cite{deFlorian:2017lwf}, and 
\texttt{NNFF}~\cite{Bertone:2017tyb} have also investigated 
$K^0_S$ FFs. Details of these analyses are discussed elsewhere, and 
we direct the reader to our previous paper~\cite{Soleymaninia:2020ahn} 
for a review.

In our previous investigations, referred to as {\tt SAK20}, we 
incorporated experimental data from the single inclusive 
electron-positron annihilation process (SIA) to determine 
FFs for $K^0_S$ production. We provided FF sets for $K^0_S$, along 
with their associated uncertainties determined through the Hessian 
method, at both next-to-leading order (NLO) and NNLO precision. 
Additionally, in \cite{Soleymaninia:2020ahn} we had examined the 
effect of hadron mass corrections as well.

Recently, the \texttt{BESIII} detector has performed measurements 
of the normalized differential cross sections for inclusive $K^0_S$ 
production in $e^+e^-$ annihilation, covering six energies from 2.2324 
to 3.6710~GeV with relative hadron energies, $z$, ranging from 0.2 
to 0.9 \cite{BESIII:2022zit}. Since this energy range is not 
well-covered by previous experimental data, it will provide valuable 
input for a QCD analysis to determine the kaon FFs. This new dataset  
motivates us to revisit and update our previous analysis of $K^0_S$ 
FFs~\cite{Soleymaninia:2020ahn}, now incorporating the \texttt{BESIII} 
measurements. Our new FF sets, {\tt FF24-}$K^0_S$, at both NLO and 
NNLO, are publicly available in the LHAPDF format. 

In recent years, machine learning (ML) has gained increasing popularity 
in various domains of particle physics, particularly in collider physics. 
One promising application of ML methods is their role in advancing our 
understanding of non-perturbative quantities related to nucleons, 
such as parton distribution functions (PDFs) and 
FFs~\cite{Khalek:2021gxf,Bertone:2017tyb,NNPDF:2017mvq}. 
Given this trend, we chose to leverage artificial neural networks (NN) 
for the extraction of FFs for $K^0_S$ hadrons in this work. Utilizing 
neural networks is expected to minimize the bias in the parameterization 
of the FFs. In addition, we use the Monte Carlo sampling method as a robust 
statistical approach to account for uncertainties in experimental data 
and derive probability density distributions of FFs from the data. To 
facilitate this analysis, we utilize the publicly available code 
\texttt{MontBlanc}, accessible from~\cite{MontBlanc}.

The organization of this paper is as follows:
In Sec.~\ref{experimental} we present and discuss the experimental 
observable for $K^0_S$ production in the SIA process. The theoretical 
framework employed in this work as well as the hadron mass corrections 
are presented in Sec.~\ref{Theoretical}. Sec.~\ref{optimization} deals 
with the $\chi ^2$ minimization and the methodology for the calculation 
of FFs uncertainties. Our main results from this study are presented 
and discussed in detail in Sec.~\ref{Results}. Finally, 
Sec.~\ref{Conclusion} provides our summary and conclusions.

%
\section{Experimental data}
\label{experimental}
%

This analysis relies on experimental measurements of the $K^0_S$ 
production cross sections in the SIA process. Various collaborations 
have reported SIA data for the cross section of $K^0_S$ hadrons in 
the final state, considering the longitudinal momentum fraction $z$ 
at different energy scales. The majority of differential cross sections 
have been normalized to the total cross section.

In this analysis, we incorporate a comprehensive set of experimental 
data from various collaborations conducted at \texttt{CERN}, 
\texttt{DESY}, \texttt{SLAC}, \texttt{KEK}, and \texttt{BESIII}. 
Our dataset encompasses untagged data from the \texttt{TASSO} 
collaboration at $\sqrt{s} = 14$, $22$, and 
$34$~GeV~\cite{Althoff:1984iz}, as well as at $\sqrt{s} = 14.8$, $21.5$, 
$34.5$, $35$, and $42.6$~GeV~\cite{Braunschweig:1989wg}. 
Additionally, we include measurements from the 
\texttt{HRS}~\cite{Derrick:1985wd}, 
\texttt{TPC}~\cite{Aihara:1984mk}, and 
\texttt{MARK II}~\cite{Schellman:1984yz} 
collaborations at $\sqrt{s} = 29$~GeV. 
Our dataset also includes data from the \texttt{CELLO} 
collaboration at $\sqrt{s} = 35$~GeV~\cite{Behrend:1989ae} and the 
\texttt{TOPAZ} collaboration at $\sqrt{s} = 58$~GeV~\cite{Itoh:1994kb}. 

Included in our analysis are also datasets measured at $\sqrt{s}=M_Z$ 
from \texttt{ALEPH}~\cite{Barate:1996fi}, 
\texttt{DELPHI}~\cite{Abreu:1994rg}, and 
\texttt{OPAL}~\cite{Abbiendi:1999ry}, as well as untagged and 
tagged quark data (light quarks $(u,d,s)$ and heavy quarks $c$ and $b$) 
from the \texttt{SLD}~\cite{Abe:1998zs} collaboration. 
  
The primary motivation for this analysis is the incorporation of the 
most recent measurements of normalized differential cross sections 
for inclusive $K^0_S$ production in SIA processes reported by the 
\texttt{BESIII} collaboration~\cite{BESIII:2022zit}. Their dataset 
spans six center-of-mass energies ranging from $\sqrt{s} = 2.2324$ 
to $3.6710$~GeV and features a $z$ coverage from $0.2$ to $0.9$. 
Collected with the \texttt{BESIII} detector at \texttt{BEPCII}, these 
new measurements offer crucial input for fitting FFs in the region 
with $\sqrt{s} < 10$ GeV, where precision SIA data have been scarce.
 
As Fig.~3 of the \texttt{BESIII} publication~\cite{BESIII:2022zit} 
shows, the comparison between the $K^0_S$ data and different 
theoretical predictions indicates a discrepancy, likely attributed 
to the fact that existing FFs had been determined from SIA data at 
high energies and have to be extrapolated to the lower energy scales 
of the BESIII measurement. One can conclude that BESIII data should 
be used to improve the fits of FFs at the low energy scales. 

\begin{figure}[t]
\vspace{0.20cm}
\includegraphics[clip,width=0.5\textwidth]{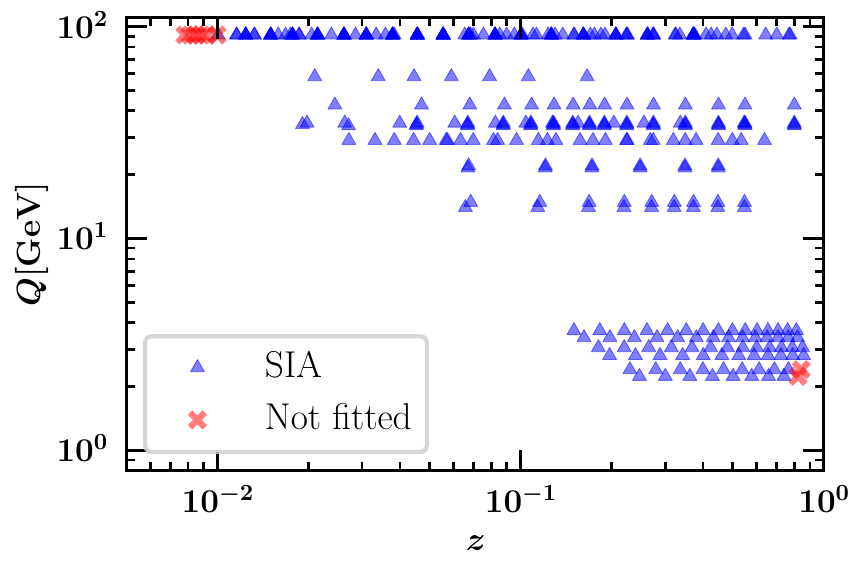}
\begin{center}
\caption{
{\small Kinematic range of the experimental SIA data in 
the ($z, Q$) plane used to determine the $K^0_S$ FFs. } 
\label{dataK0}
}
\end{center}
\end{figure}

Kinematic cuts are implemented to ensure that the selected data 
points are in a regime where perturbative fixed-order predictions 
hold reliable. The kinematic coverage of the datasets applied in this 
analysis is shown in Fig.\ref{dataK0}. We adjust the minimal value 
of $z$ at $z_{\rm min} = 0.013$ for all experiments. For most of the 
experiments, the upper limit $z_{\rm max}$ is set at $0.9$, except 
for two datasets from \texttt{BESIII}: the data at the energies 
$\sqrt{s} = 2.2324$ and $2.4$~GeV are cut at $z_{\rm max} = 0.8$. 
Due to outliers observed in the final data points of these datasets,
 their inclusion in the fit leads to an increased $\chi^2$ value without
  substantially altering the resulting FFs.
   Hence, there is no compelling rationale for integrating them into the analysis.
Overall, we omit just 8 data points across all the SIA datasets 
used in this analysis and find that $N_{\rm dat} = 346$ data points 
can be used.

It is worth mentioning that, unlike in our previous analysis 
\cite{Soleymaninia:2020ahn}, we chose not to incorporate the 
experimental data from \texttt{HRS} \cite{Derrick:1985wd} and 
\texttt{DELPHI} at 183 and 189~GeV \cite{Abreu:2000gw} in this study. 
This decision was made since we could not find a satisfactory 
fit for these particular datasets. Specifically, their inclusion 
resulted in a noticeable decline in the overall quality of the fit, 
with $\chi ^2$ values per data point exceeding 7 for \texttt{HRS}, 
approximately 7 for \texttt{DELPHI} at 183~GeV, and surpassing 17 
at 189~GeV. This discrepancy is evident when comparing the high 
$\chi ^2$ values per data point observed in our previous analysis 
and in {\tt AKK08}~\cite{Albino:2008fy} to the even poorer performance 
in the current analysis. We have identified the source of this 
discrepancy as an apparent inconsistency between these three 
datasets and the others, particularly with BESIII.

%
\section{Theoretical setup}
\label{Theoretical}
%

The theoretical setup of our analysis closely follows that illustrated 
in Sec.~III of Ref.~\cite{Soleymaninia:2020ahn}. In this section, we 
recapitulate the theoretical understanding of standard collinear 
factorization, comprising perturbative and non-perturbative components 
of the cross-section for the SIA process. We also review the 
time-like DGLAP evolution of FFs, alongside detailing numerical 
computations of SIA cross-sections, including theoretical settings 
and physical parameters. 

The total cross-section for $e^+ e^-$ annihilation in the production 
of hadrons ($e^+ e^- \rightarrow h+X$) can be expressed in terms of 
the convolution of coefficient functions and FFs
\begin{eqnarray}
\label{cross-section}
\frac{1}{\sigma_{tot}}
\frac{d\sigma^h} {dz} &=&
\frac{1}{\sigma_{tot}} 
C_{i}\left(x, \alpha_{s}(\mu), \frac{Q^2}{\mu ^2}\right) 
\otimes
D_i^h\left(\frac{z}{x}, \mu ^2 \right) 
\,.
\end{eqnarray}
The coefficient functions $C_{i}(x, \alpha_{s} (\mu), Q^2 / \mu ^2)$ 
are currently computed up to $\mathcal{O}(\alpha_{s}^2)$ in 
perturbative QCD \cite{Mitov:2006wy,Mitov:2006ic} and 
$D_i^h(z / x, \mu^2)$ denotes the FFs. The parton FFs at different 
energy scales are connected through the time-like DGLAP evolution 
equations~\cite{Dokshitzer:1977sg}, given by:
\begin{eqnarray} 
\label{DGLAP}
\frac{\partial D^h_i(z, \mu^2)}
{\partial \ln \mu^2} 
= P_{ji}\left(x, \alpha_s(\mu^2)\right) 
\otimes
D_j^h\left(\frac{z}{x}, \mu^2 \right) 
\,.
\end{eqnarray}
In this equation, $P_{ji} (x, \alpha_s(\mu^2))$ represent the 
time-like splitting functions and $\otimes$ denotes the convolution integral.

When incorporating the mass of the produced hadrons into the kinematic 
calculations, hadron mass corrections modify the FFs. In the context 
of the SIA process, these corrections are very large and therefore 
indispensable, in particular at low values of $z$. Hadron mass effects 
are introduced by employing light-cone coordinates and can be taken 
into account by using a modified scaling variable $\eta$ instead of $z$: 
\begin{eqnarray}
\label{eta_mass}
\eta=
\frac{z}{2}
\left(1 + \sqrt {1-
\frac{4 m_{h}^{2}}
{s z^{2}}}\right),
\end{eqnarray}
where $m_h$ is the hadron mass. With this change of variables 
we can express the cross section as 
\begin{eqnarray}
\label{massive}
\frac{d\sigma} {dz}=
\frac{1}
{1 - {m_{h}^{2}}/
	{s \eta^{2}}} \sum_a \int_\eta^1 \frac{dx_a}{x_a}
\frac{d\hat{\sigma}_{a}}{dx_{a}}
D_{a}^{h} \left(\frac{\eta}{x_{a}}, \mu \right) 
,
\end{eqnarray}
Here, $d\hat{\sigma}_{a} / dx_{a}$ denotes the differential 
cross section of the hard sub-process with parton $a$, calculated 
in pQCD, and $a$ runs over all quark flavors and the gluon.  
We use Eq.~(\ref{massive}) for the theory predictions of this work.
In this study, we compute the $K^0_S$ hadron FFs within the framework 
of the Zero-Mass Variable-Flavor-Number Scheme (ZM-VFNS), where all 
active flavors are treated as massless. However, the inclusion of 
heavy quark masses is necessary in order to calculate the number 
of active flavors taking into account the correct heavy-quark 
thresholds. 

In this analysis, fixed values are adopted for the charm and bottom 
quark masses, setting $m_c$ = 1.51~GeV and $m_b$ = 4.92~GeV, 
respectively. The strong running coupling is calculated at two-loop 
order by setting $\alpha_{s}(m_Z = 91.1876~\mathrm{GeV}) = 0.118$.

\begin{figure}[h!]
\vspace{0.20cm}
\includegraphics[clip,width=0.50\textwidth]{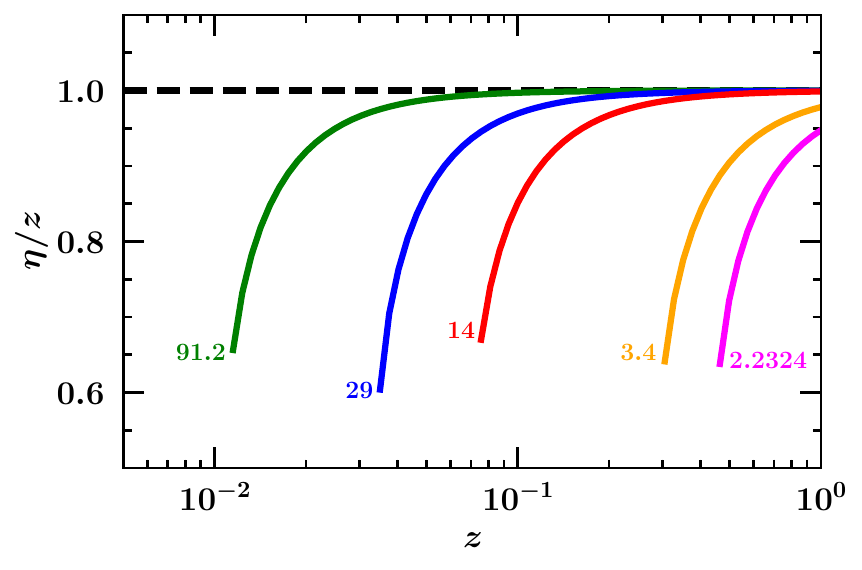}
\begin{center}
\caption{
{\small Ratio $\eta/z$ as a function of $z$ at five representative 
values of $\sqrt{s}$. } 
\label{eta_vs_z}
}
\end{center}
\end{figure}

To illustrate the need for hadron mass corrections, we present 
in Fig.~\ref{eta_vs_z} the ratio $\eta/z$ as a function of $z$ at five 
representative values of $\sqrt{s}$. This figure indicates that 
the deviation of the momentum fraction available to the hadron 
($\eta$) from $z$ becomes significant when $z$ and/or $\sqrt{s}$ 
decrease. We see that at $\sqrt{s}=2.2324$~GeV and $\sqrt{s}=3.4$~GeV 
this correction is sizeable in the full range of $z$ values. 
The corrections 
are larger than $20\%$ for $z<0.6$ at $\sqrt{s}=2.2324$~GeV, $z<0.3$ 
at $\sqrt{s}=3.4$~GeV, $z<0.1$ at $\sqrt{s}=14$~GeV, $z<0.04$ at 
$\sqrt{s}=29$~GeV, and $z<0.015$ at $\sqrt{s}=91.2$~GeV. Corrections 
exceeding $50\%$ can be observed around $z<0.4$ at $\sqrt{s}=2.2324$~GeV, 
$z<0.3$ at $\sqrt{s}=3.44$~GeV, $z<0.06$ at $\sqrt{s}=14$~GeV, 
$z<0.03$ at $\sqrt{s}=29$ GeV, and $z<0.01$ at $\sqrt{s}=91.2$~GeV. 
Since we include the new \texttt{BESIII} data, which have been 
measured from $2.2324$ to $3.6710$~GeV, the hadron mass corrections 
are essential in the kinematic region covered by the data and 
significantly improves the description of the data used in this 
analysis. 

For the cross sections, Eq.~(\ref{massive}), our investigations show 
that at the energy scale of $91.2$ GeV, the hadron mass correction 
is approximately $\pm 4\,\%$. At lower values of $\sqrt{s}=29$ GeV, 
relevant for the TPC data, this correction varies approximately 
in the range between $\sim-20\,\%$ and $\sim+6\,\%$ depending on the 
value of $z$, while for the case of the TASSO 14 data, the correction 
is between $-20\,\%$ and $+12\,\%$. Our finding is that for energies of 
$3.4$~GeV and $2.2324$~GeV, mass corrections become more important 
and reach $\sim\pm45\,\%$ at $\sqrt{s}=3.4$~GeV and $\sim+35\,\%$ 
to $\sim-50\,\%$ at $\sqrt{s}=2.2324$~GeV for BESIII. All in all, 
hadron mass corrections prove to be significant in mid-range 
energies and indispensable for low-energy experiments conducted 
at BESIII.

As outlined in our previous publication~\cite{Soleymaninia:2020ahn}, 
the analysis of inclusive SIA data only allows one to determine the 
combined quark and antiquark FFs, i.e.\ for the quark combinations 
$q^+ = q + \bar{q}$. SIA cross section data are available for tagged 
light, charm, and bottom quarks. Therefore we can adopt a flavor 
decomposition approach and, keeping in mind the flavor content of 
$K_S^0$, we chose the following combinations to be fitted: 
\begin{eqnarray}
D^{K_S^0}_{u^+}, 
~D^{K_S^0}_{d^+}, 
~D^{K_S^0}_{s^+}, 
~D^{K_S^0}_{c^+}, 
~D^{K_S^0}_{b^+}, 
~D^{K_S^0}_g.
\label{eq:SIAcombinations}
\end{eqnarray}
In the next section we describe the optimization methodology and 
the method of FFs determination.

%
\section{Optimization methodology and uncertainty}
\label{optimization}
%

The determination of unknown parameters or functions typically 
involves employing a maximum log-likelihood method. Assuming small 
uncertainties and therefore an approximately linear parameter dependence, 
the problem can be reduced to minimizing an appropriately chosen 
$\chi^2$. In this work we use neural networks to describe the 
theoretical predictions and choose to minimize the following $\chi^2$ 
function: 
\begin{equation}
\label{eq:chi2}
\chi^{2(k)} \equiv  
\left(\boldsymbol {\rm T}(\boldsymbol 
{\theta}^{(k)})- 
\boldsymbol {x}^{(k)}\right)^{\rm T} 
\cdot \boldsymbol {\rm C}^{-1}\cdot 
\left(\boldsymbol{\rm T}
(\boldsymbol {\theta}^{(k)})- 
\boldsymbol {x}^{(k)}\right) 
\, .
\end{equation}
Here, $k$ is the replica number, $\boldsymbol {x}^{(k)}$ denotes 
pseudo-data for replica $k$ which are obtained by randomly shifting 
the central values proportional to their uncertainties, and 
$\boldsymbol {\theta}$ represents the set of weights and biases in 
the neural network, $\boldsymbol{\rm C}$ represents the covariance 
matrix including all uncertainties and (possible) correlations of the 
measured data $\boldsymbol{x}$, and finally $\boldsymbol {\rm T}$ 
represents the theoretical prediction. We use the open source framework 
\texttt{MontBlanc} in all the analyses performed in this work. In this 
framework, minimization of $\chi^2$ is performed by the \texttt{Ceres 
Solver} open source code~\cite{ceres-solver}. The NN parameterization 
is provided by the \texttt{NNAD} code~\cite{nnad}. The core of our 
theoretical predictions relies on the \texttt{APFEL++} 
package~\cite{Bertone:2017gds}. Using \texttt{NNAD} an analytic 
expression for $\chi^2$ is formed and \texttt{Ceres Solver} is used 
to solve the corresponding minimization problem using trust region 
methods. The FFs for each parton flavor are expressed through a NN 
model at the initial scale $Q_0 = 5$~GeV as: 
\begin{eqnarray}
\label{NN}
zD^{h^+}_i(z,Q_0)
= 
(N_i(z; \boldsymbol {\theta}) - N_i(1; \boldsymbol {\theta}))^2 
\, .
\end{eqnarray}

$N_i(z;\boldsymbol\theta)$ denotes the output generated by the neural network. We opt not to incorporate pre-factors like $z^\alpha(1-z)^\beta$ to regulate the behavior at low and high values of $z$. However, it is important to note that the NN output is subtracted at $z=1$, ensuring that the fragmentation functions vanish at this specific kinematic point. Moreover, the resulting output is squared to guarantee that FFs remain non-negative.

The neural network is a type of multi-layer feed-forward perceptron~\cite{Forte:2002fg}. It comprises nodes organized into sequential layers, where the output $\xi_i^{(l)}$ of the $i$-th node in the $l$-th layer is,
\begin{equation}
  \xi_i^{(l)}=g\left(\sum_j\omega_{ij}^{(l)}\xi_j^{(l-1)}+\theta_i^{(l)} \right)
  \,\mbox{.}
\label{eq:actfunc}
\end{equation}
The function $g$ is called \textit{activation function}, which is a sigmoid function for hidden layers and linear for the output layer. 
The parameters
$\left\{\omega_{ij}^{(l)},\theta_i^{(l)}\right\}$, denote {\it weights} and {\it biases} respectively, and are determined by 
minimization of the $\chi^2$ function.

We use a simple but effective method in this study. Its effectiveness 
is based on the so-called universal approximation 
theorem~\cite{Csaji:2001}, 
which states that a simple feed-forward NN like the one we are using 
can represent any function in any specified range. Our NN setup has 
just one hidden layer, and we pick 25 nodes for it. We have one node 
in the input layer, which is the value of $z$, and 6 output nodes, one 
for each combination of FFs in Eq.~(\ref{eq:SIAcombinations}). This 
architecture is therefore denoted as (1-25-6) and has 175 weight and 
31 bias parameters, i.e.\ 206 parameters in total to be fitted 
to the data. Fewer nodes could work just as well. We have tried a 
NN setup with two hidden layers, (1-9-9-6), to see if this would 
change our results. Unsurprisingly, our tests showed that the 
results are consistent, as other research supports this 
idea~\cite{Khalek:2021gon}. 

We use cross-validation to avoid overfitting our FFs. For each 
data replica, the datasets with more than 10 points are randomly 
split into training and validation subsets, each containing half 
of the points. We only use the training group to perform the fit. 
If there are 10 or fewer points, we use all of them for training. 
During the training step, the $\chi^2$-value of the validation 
set is monitored. The fit is stopped when $\chi^2$ of the 
validation set reaches its minimum. Replicas whose total $\chi^2$ 
per point is larger than three are discarded.
We are therefore convinced that our results do indeed describe the 
data, rather than the specific setup of the NN. Finally, we decided 
to use 100 replicas for our analysis, as it was shown in a previous 
study to be enough to represent the central values and uncertainties 
of the FFs~\cite{Soleymaninia:2022qjf}. 

The Monte Carlo approach has become a prevalent technique in various 
QCD analyses for error propagation, as evidenced by its application 
in studies such as~\cite{Sato:2016tuz,Sato:2016wqj, 
Moutarde:2019tqa,NNPDF:2017mvq,AbdulKhalek:2019mzd}. 
Here, this method involves estimating the posterior probability 
distribution of NN parameters through fitting multiple times. Each fit 
is conducted independently using a pseudo dataset, known as a replica, 
resulting in an optimal set of parameters and therefore FFs. 
Subsequently, the results of all fits collectively encapsulate the 
probability distribution of the data in FFs. Therefore, we define 
both the central value as the mean of the replicas, and the 
uncertainties of FFs as the standard deviation over the replica set.

%
\section{Results and discussion} 
\label{Results}
%

The main results and findings of this study are presented 
and discussed in detail in this section. First, we 
evaluate the quality of the \texttt{FF24-}$K^0_S$ fit 
by examining the total and individual values of $\chi^2$ per dataset. 
Second, we perform a comparison between the analyzed 
experimental data and the predictions obtained using our $K^0_S$ 
FFs for all the datasets included in this analysis. 
The kinematical cuts are also discussed in this section. 
Finally, we compare our FFs with other determinations 
available in the literature and illustrate the 
origin of some specific features of our new FFs.

\begin{table*}[htb]
\renewcommand{\arraystretch}{2}
\centering 	\scriptsize
\begin{tabular}{lccccccr}				\hline
Experiment   & Reference & ~ $\sqrt{s}$ GeV ~&~ $z_{min}$ ~&~ $z_{max}$ ~&~ $\chi^{2}_{\tt NLO}/\#\mathrm{data}$ ~&~ $\chi^2_{\tt NNLO}/\#\mathrm{data}$    \\
				\hline \hline
				{BESIII} &\cite{BESIII:2022zit} &  2.2324 & 0.013 & 0.8 & 0.92&0.94\\
				{BESIII}  &\cite{BESIII:2022zit} &  2.4 & 0.013 & 0.8& 0.14&0.20\\
				{BESIII} &\cite{BESIII:2022zit} &  2.8 & 0.013 & 0.9& 2.55&2.18\\
				{BESIII}  &\cite{BESIII:2022zit} &  3.05 & 0.013 & 0.9& 1.37&1.39\\
				{BESIII} &\cite{BESIII:2022zit} &  3.4 & 0.013& 0.9& 0.84&0.85\\
				{BESIII} &\cite{BESIII:2022zit} &  3.6710 & 0.013 & 0.9& 0.55&0.56\\		
				{TASSO}  &\cite{Althoff:1984iz}   &  14  & 0.013& 0.9& 0.68&0.63\\
				{TASSO}  &\cite{Braunschweig:1989wg}    &    14.8 & 0.013& 0.9& 1.60&1.55\\
				{TASSO}  &\cite{Braunschweig:1989wg}  & 21.5 & 0.013  & 0.9&1.19&1.07\\
				{TASSO}  &\cite{Althoff:1984iz}  &  22 & 0.013 & 0.9&   1.27&1.17\\       			
				{TPC}    &\cite{Aihara:1984mk}   &  29 & 0.013& 0.9&  0.23& 0.22\\
				{MARKII} &\cite{Schellman:1984yz}  & 29 & 0.013& 0.9  &  0.37& 0.33\\
				{TASSO}  &\cite{Althoff:1984iz}  &  34 & 0.013 & 0.9& 2.04& 1.83\\
				{TASSO}  &\cite{Braunschweig:1989wg} &  34.5 & 0.013& 0.9&    1.27& 1.21  \\
				{ TASSO} &\cite{Braunschweig:1989wg}&  35 & 0.013 & 0.9&    0.92&  0.85\\
				{CELLO}  &\cite{Behrend:1989ae}&   35 & 0.013& 0.9& 0.40& 0.35\\
				{TASSO}  &\cite{Braunschweig:1989wg}  &  42.6 & 0.013 & 0.9 &  1.19& 1.15\\
				{TOPAZ}  &\cite{Itoh:1994kb}   &  58& 0.013 & 0.9& 0.34&0.32 \\
				{ALEPH}  &\cite{Barate:1996fi}  & 91.2 & 0.013 & 0.9 &  0.35& 0.32\\
				{DELPHI} &\cite{Abreu:1994rg}  & 91.2  & 0.013 & 0.9&  0.70&0.73\\
				{OPAL}   &\cite{Abbiendi:1999ry} &  91.2 & 0.013 & 0.9&   0.90& 0.88\\
				{SLD} total &\cite{Abe:1998zs}   &  91.2 & 0.013  & 0.9& 0.93& 0.93 \\
				{SLD} uds &\cite{Abe:1998zs}  & 91.2 & 0.013  & 0.9&  0.65&  0.73\\
				{SLD} charm &\cite{Abe:1998zs}  &  91.2 & 0.013  & 0.9 &  0.64& 0.66\\
				{SLD} bottom &\cite{Abe:1998zs}  &  91.2 & 0.013 & 0.9&  1.64& 1.46 \\
\hline \hline
 &  &  &&Total $\#$data & &  \\
\hline
Total $\chi^2/{\#\rm data }$ &  &  & & 346 & 0.91&0.87  \\
\hline \hline	
\end{tabular}
\caption{\small 
The list of input datasets included in the analysis of $K^0_{S}$ FFs 
at NLO and NNLO accuracy. For each dataset, we provide the name of the 
experiment, the corresponding published reference, the 
center-of-mass energy $\sqrt{s}$, and the value of $\chi^2$ per 
data point for the individual dataset at NLO and NNLO accuracy. 
The total value of $\chi^2$ per data point is shown in the last line. 
}
\label{tab:datasetsK0s}
\end{table*}

The experimental data used in this analysis are collected in 
Table~\ref{tab:datasetsK0s}. The table includes the reference for 
each dataset and the associated energy scale. Additionally, details 
about the kinematic cuts for both small and large regions of $z$ are 
provided. The last two columns show the $\chi ^2$ values per data 
point for each dataset at both NLO and NNLO accuracy. The table also 
presents the total $\chi ^2$ values for all the SIA datasets.

The analysis includes a total of 346 data points after the application 
of the kinematical cuts described above. The overall $\chi^2$ per data 
point is $0.91$ for the NLO and $0.87$ for the NNLO analysis, showcasing 
a highly effective description of the entire dataset for both 
perturbative orders.

A detailed examination of the entries in Table~\ref{tab:datasetsK0s} 
reveals a consistent fit quality and satisfactory description for 
each individual dataset. It is noteworthy that the incorporation of 
NNLO corrections results in a reduction of $\chi^2$ per data point 
for most of the datasets, contributing to a further enhancement of 
the overall fitting quality indicated by the total $\chi^2$.

In the following, we compare the theoretical predictions calculated 
using the \texttt{FF24-}$K^0_S$ FFs with the analyzed experimental 
data for all the datasets, focusing on NNLO accuracy.
In Figs.~\ref{fig:data1}, \ref{fig:data2}, and \ref{fig:data3}, we 
present the predictions for the normalized differential cross section 
at NNLO accuracy as a function of $z$, compared with the SIA data. 
The upper panels of the figures show the actual cross sections from 
experiment (black points) and theory (red points). In the lower 
panels of Figs.~\ref{fig:data1}-\ref{fig:data3}, the black points 
show the ratios of data over theory for each dataset, allowing for 
a better comparison of the central values of our theoretical 
predictions at NNLO accuracy with experimental data. The error 
bars of the black (red) points are obtained by normalizing the 
experimental (theoretical) uncertainties to the central values 
of our theory prediction.

{
\begin{figure*}[htb]
	\vspace{0.50cm}
	\centering
	\subfloat{\includegraphics[width=0.33\textwidth]{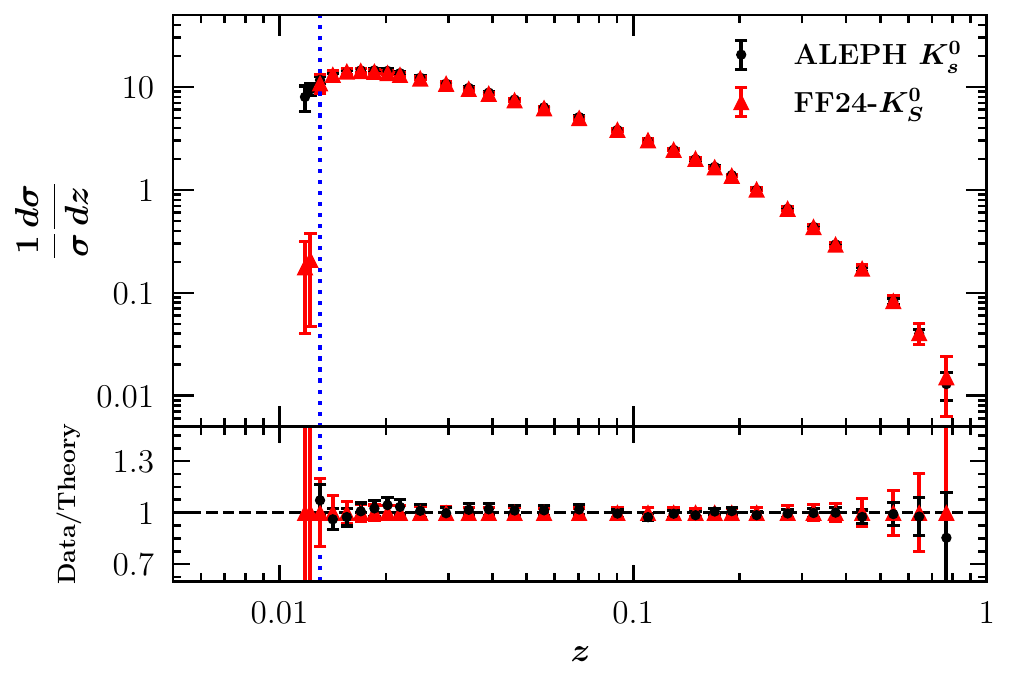}} 	
	\subfloat{\includegraphics[width=0.33\textwidth]{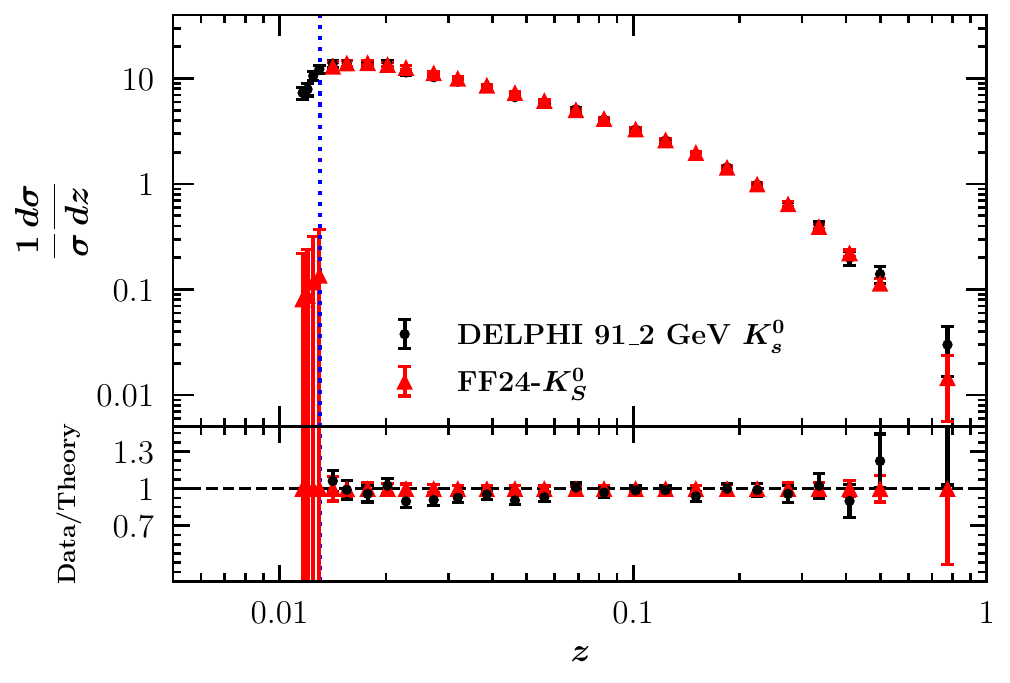}}
	\subfloat{\includegraphics[width=0.33\textwidth]{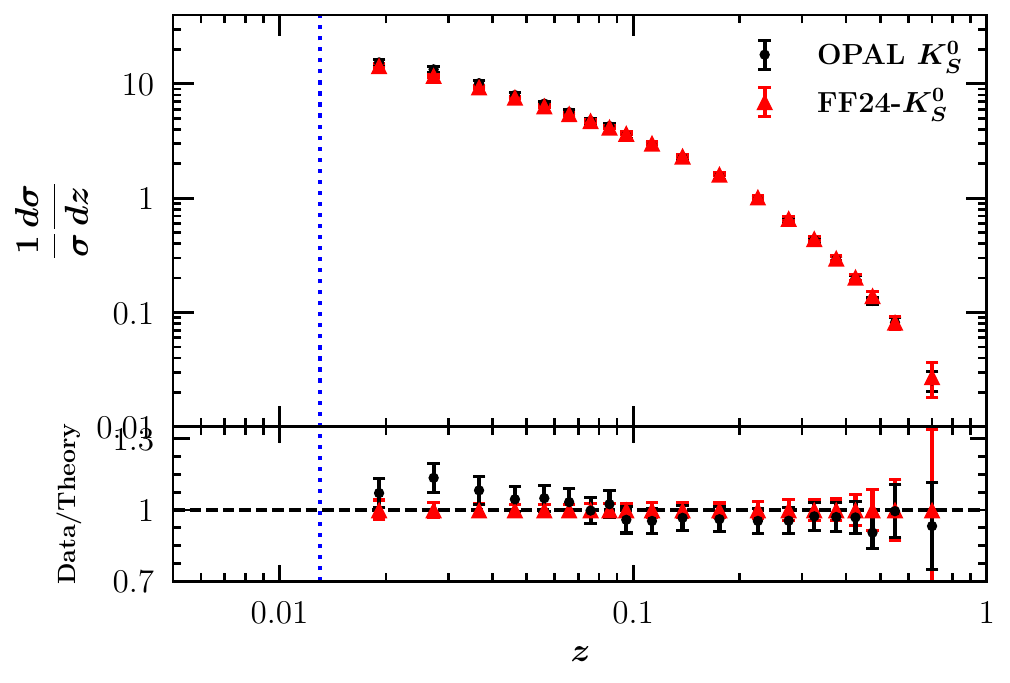}}\\	
	\subfloat{\includegraphics[width=0.33\textwidth]{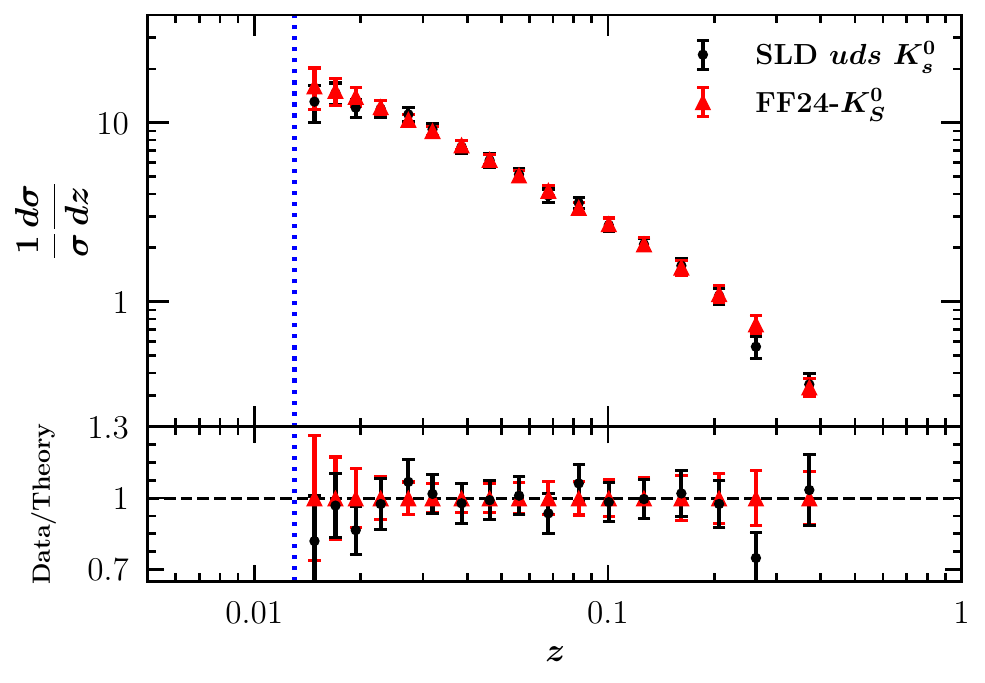}} 
	\subfloat{\includegraphics[width=0.33\textwidth]{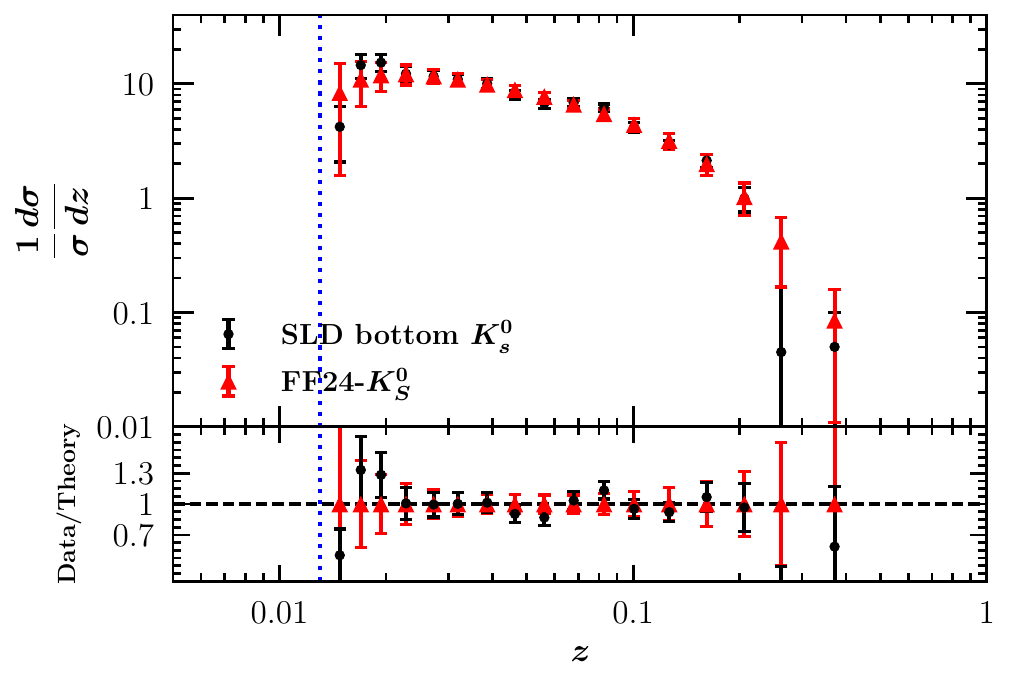}}	
	\subfloat{\includegraphics[width=0.33\textwidth]{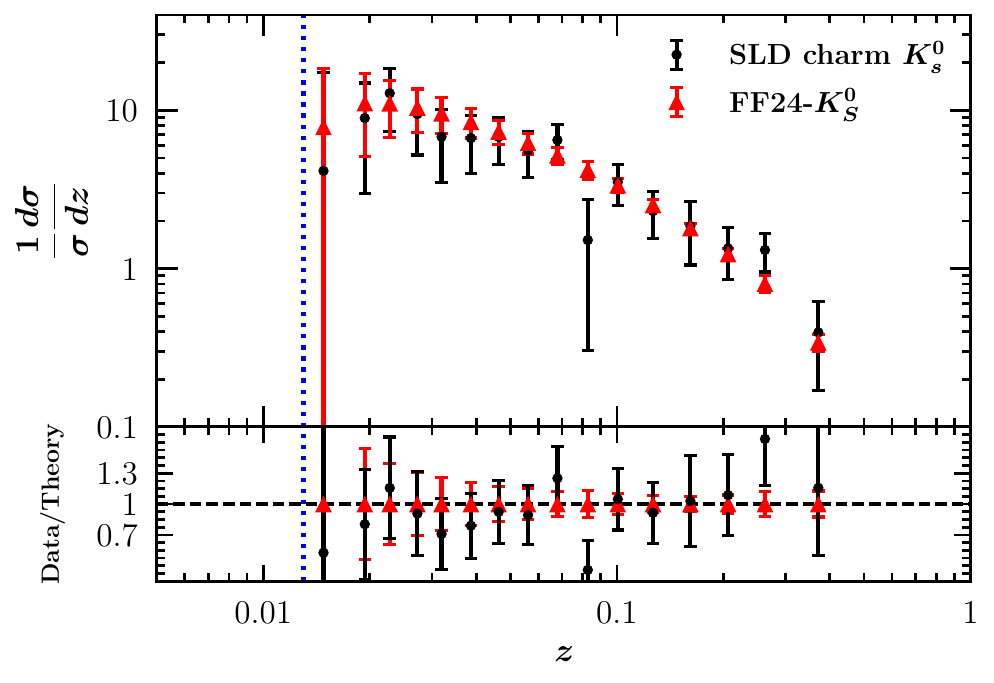}}\\
	\subfloat{\includegraphics[width=0.33\textwidth]{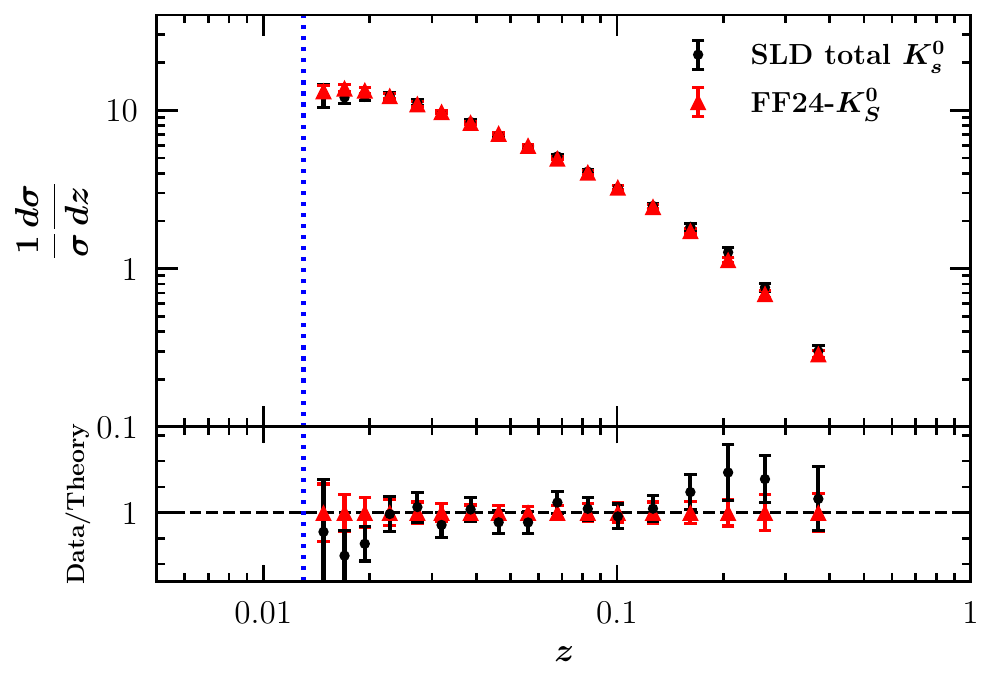}} 	
	\subfloat{\includegraphics[width=0.33\textwidth]{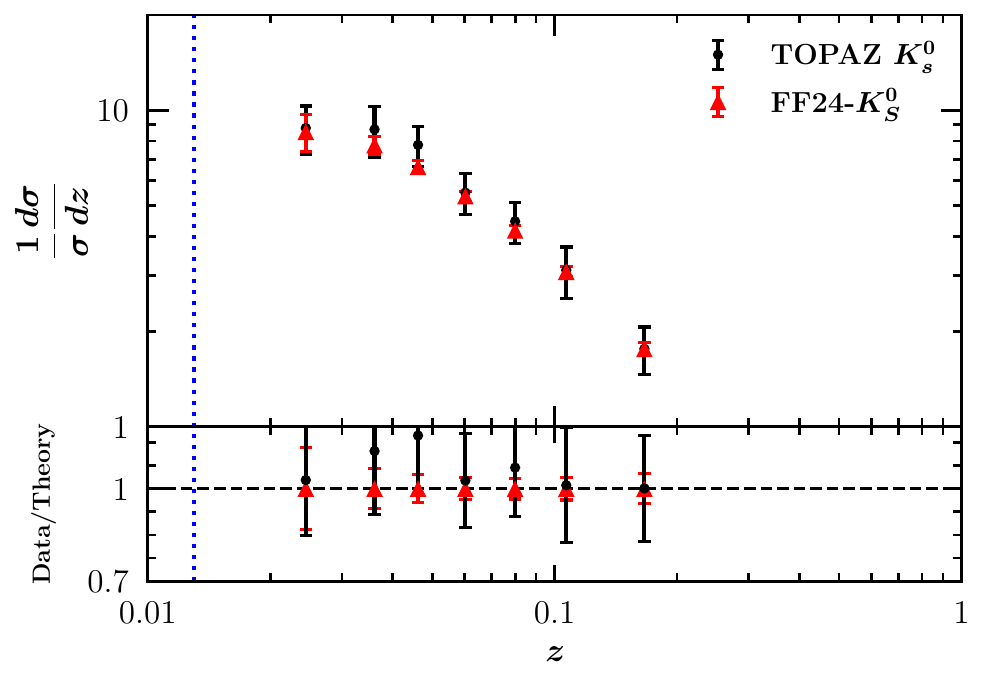}}
	\subfloat{\includegraphics[width=0.33\textwidth]{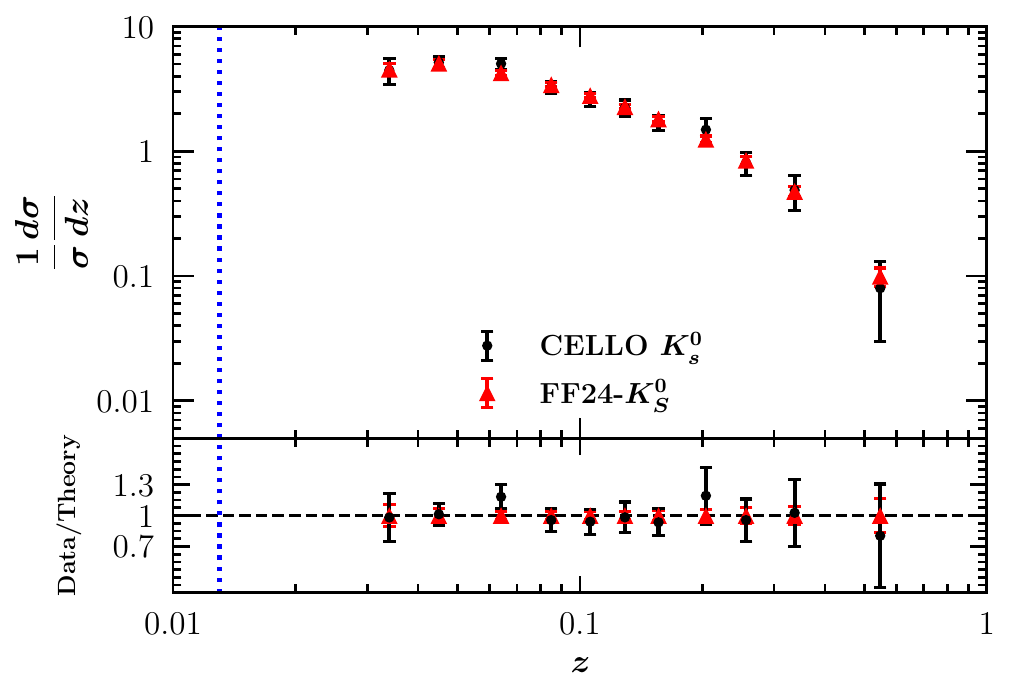}}\\		
\begin{center}
\caption{ \small 
Comparison between the $K^0_S$ production cross section calculated 
at NNLO accuracy and experimental data from different experiments, 
including {\tt ALEPH}, {\tt DELPHI91.2}, {\tt OPAL}, {\tt SLD$_{uds}$}, 
{\tt SLD$_{b}$}, {\tt SLD$_{c}$}, {\tt SLD$_{\rm total}$}, {\tt TOPAZ}, 
and {\tt CELLO}. The lower panels show the data/theory ratios.
}
\label{fig:data1}
\end{center}
\end{figure*}

\begin{figure*}[htb]
	\vspace{0.50cm}
	\centering
	\subfloat{\includegraphics[width=0.33\textwidth]{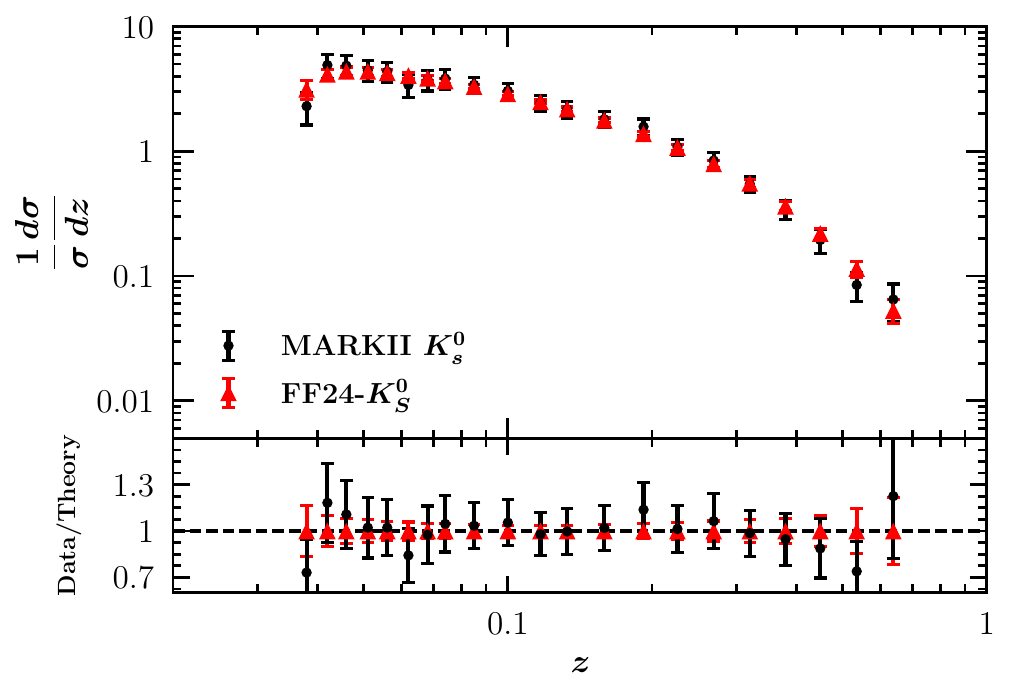}} 
	\subfloat{\includegraphics[width=0.33\textwidth]{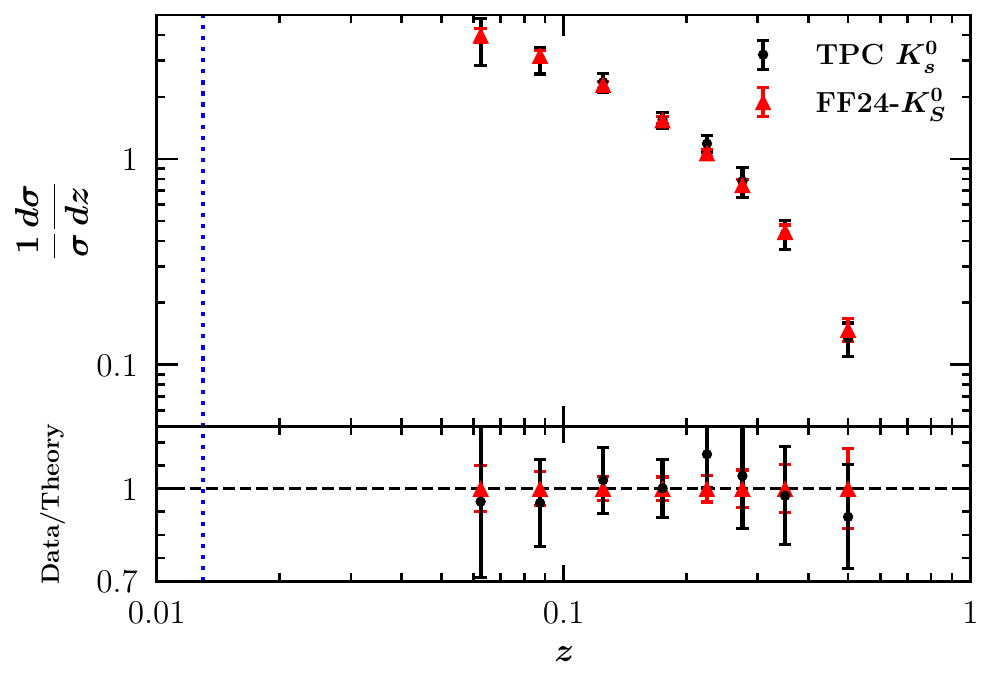}}	
	\subfloat{\includegraphics[width=0.33\textwidth]{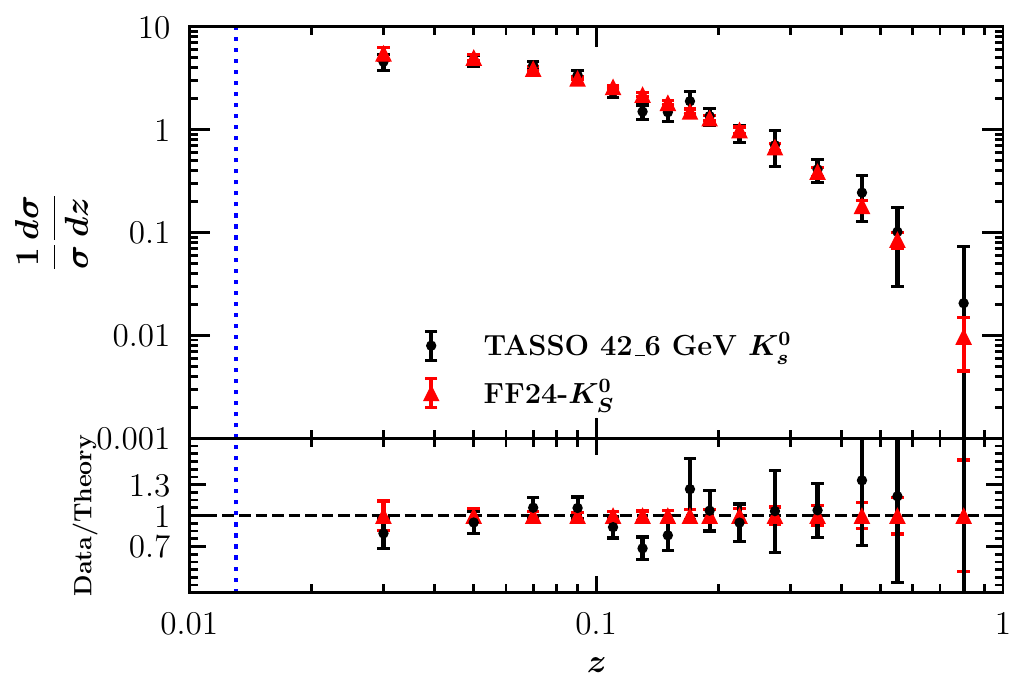}}\\	                      
	\subfloat{\includegraphics[width=0.33\textwidth]{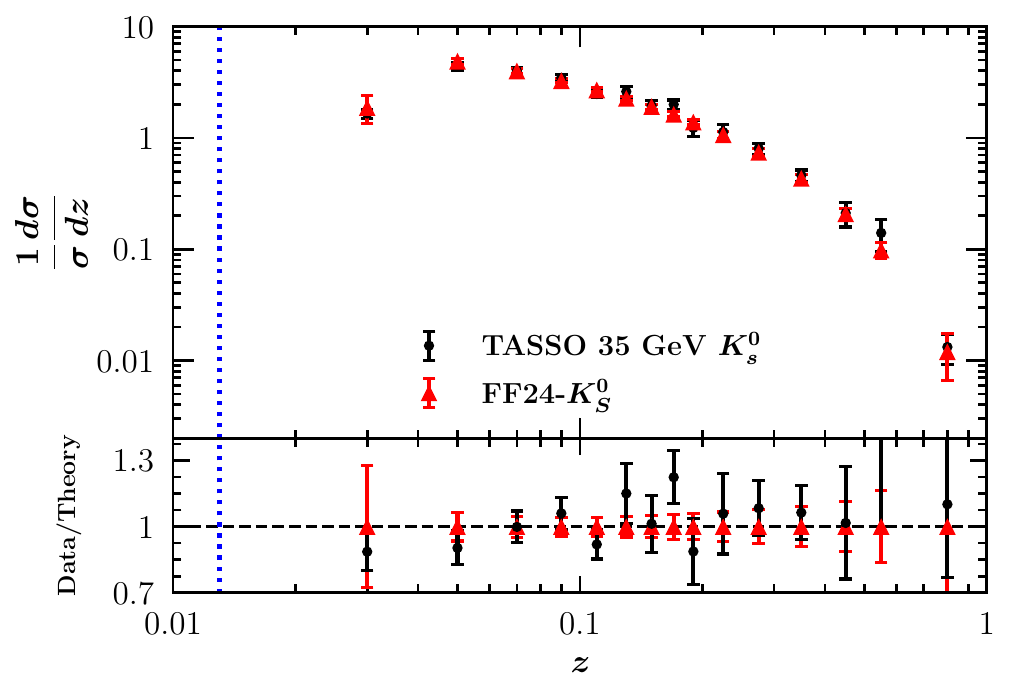}}
	\subfloat{\includegraphics[width=0.33\textwidth]{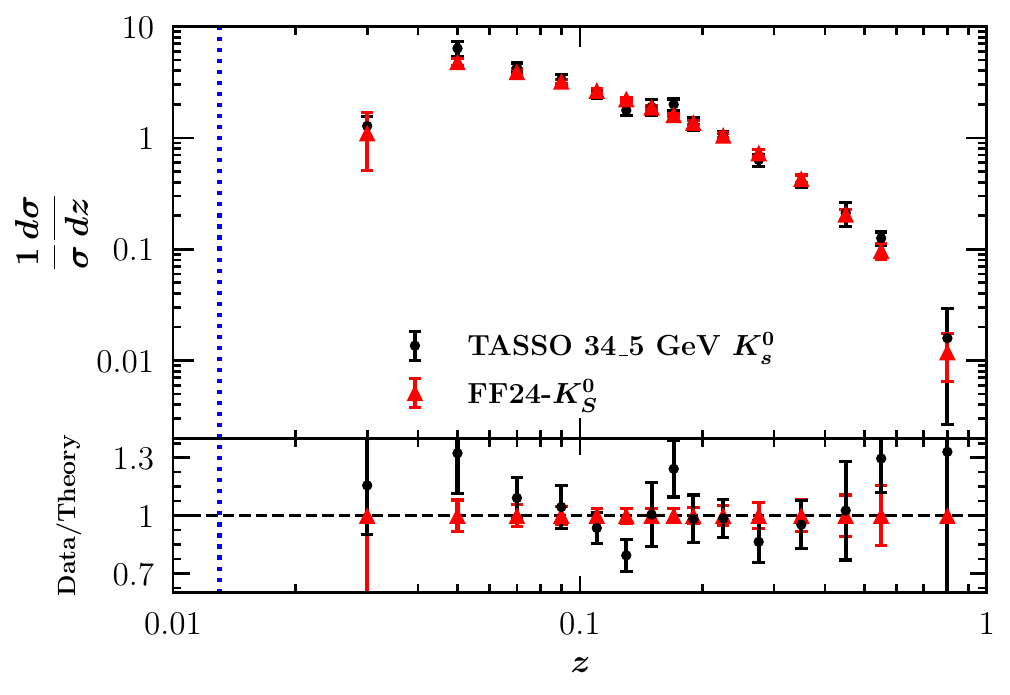}}
	\subfloat{\includegraphics[width=0.33\textwidth]{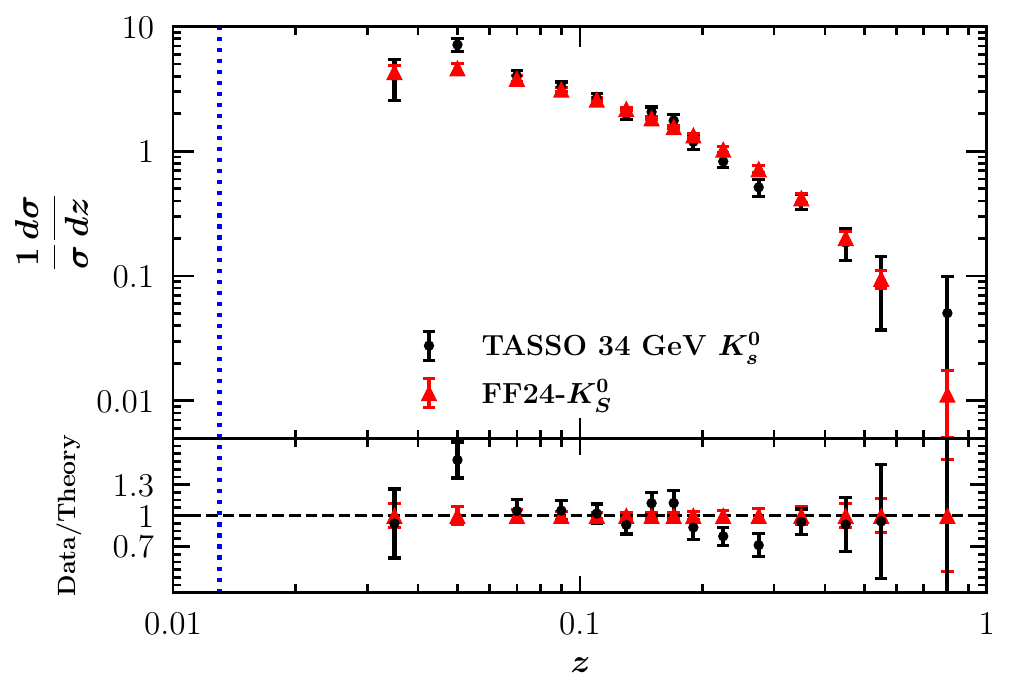}} \\
	\subfloat{\includegraphics[width=0.33\textwidth]{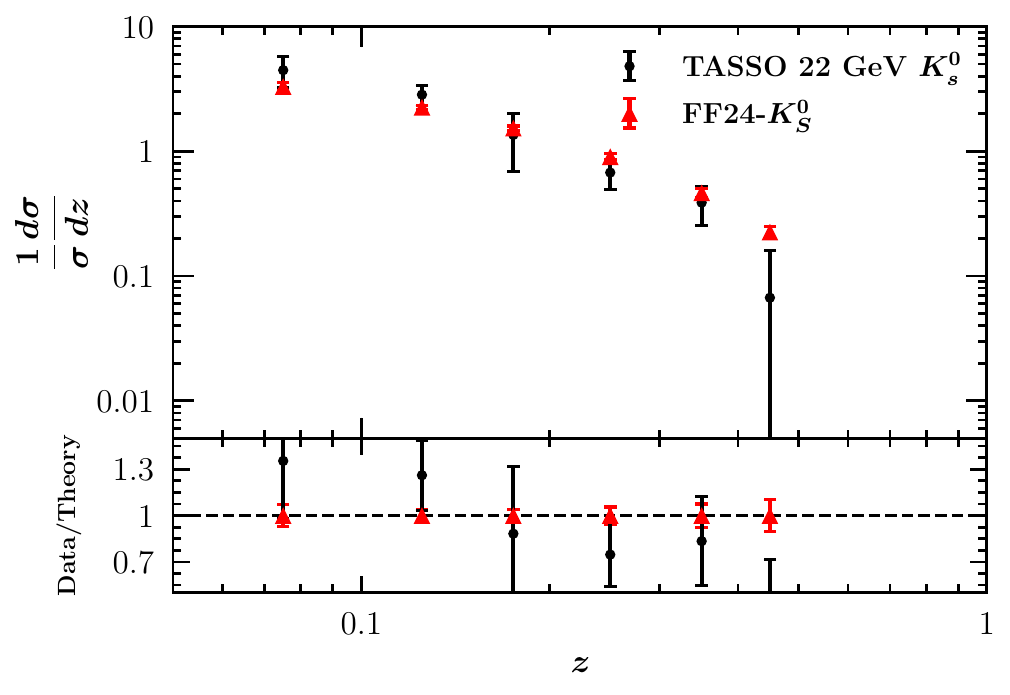}}	
	\subfloat{\includegraphics[width=0.33\textwidth]{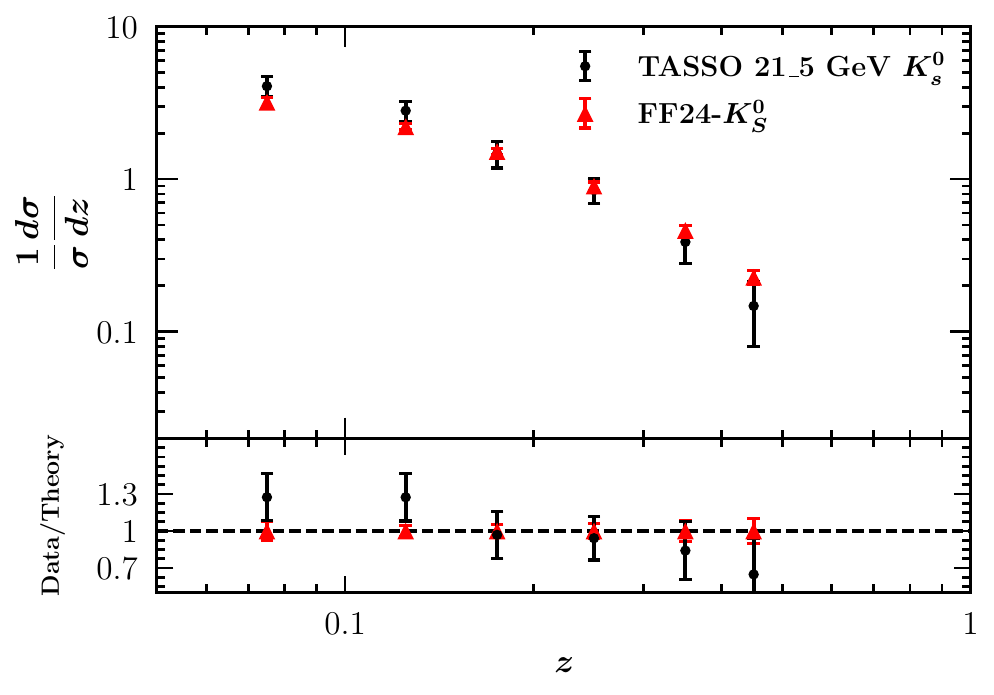}}\\	
\begin{center}
\caption{ \small 
Same as Fig.~\ref{fig:data1} but this time for some other selected experimental data, 
including {\tt MARKII}, {\tt TPC}, {\tt TASSO42.6}, {\tt TASSO35}, {\tt TASSO34.5}, {\tt TASSO34}, {\tt TASSO22}, {\tt TASSO21.5}.   }
\label{fig:data2}
\end{center}
\end{figure*}

\begin{figure*}[htb]
	\vspace{0.50cm}
	\centering
	\subfloat{\includegraphics[width=0.33\textwidth]{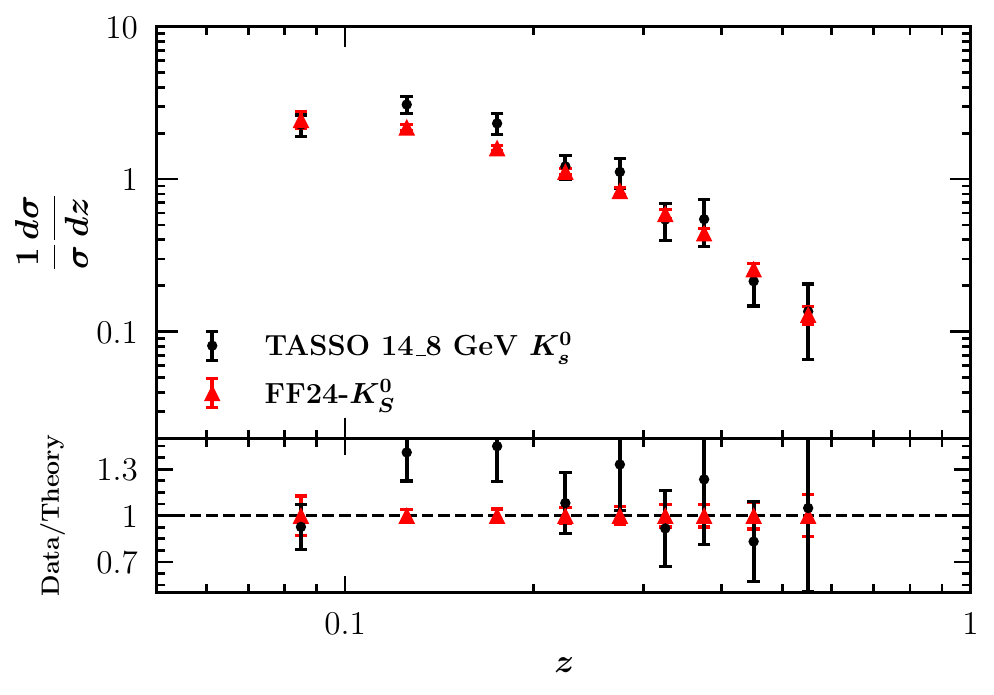}}
	\subfloat{\includegraphics[width=0.33\textwidth]{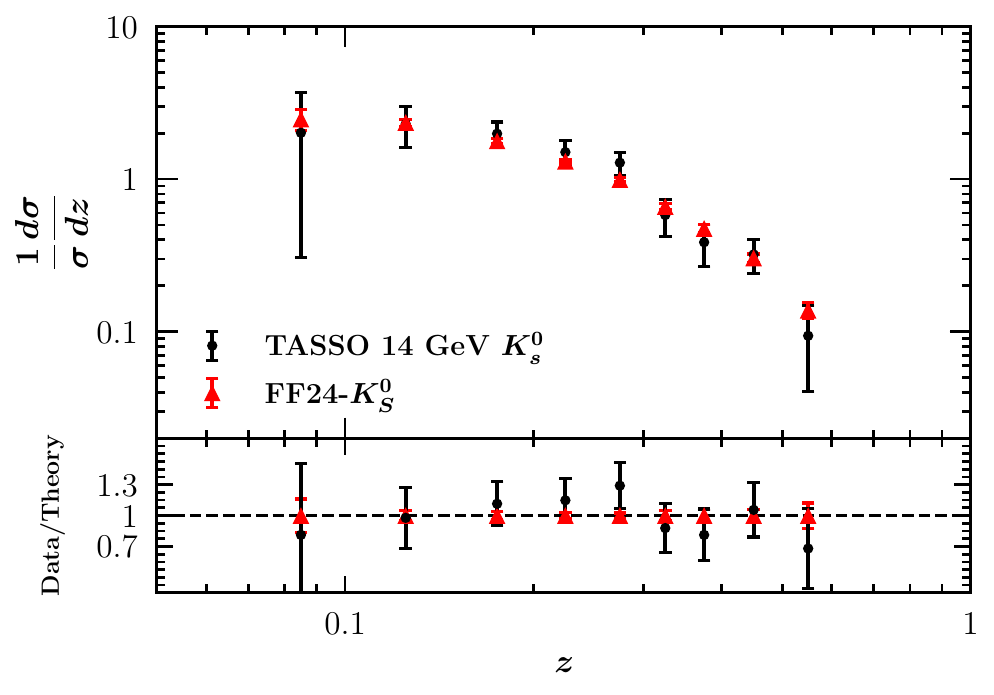}}
	\subfloat{\includegraphics[width=0.33\textwidth]{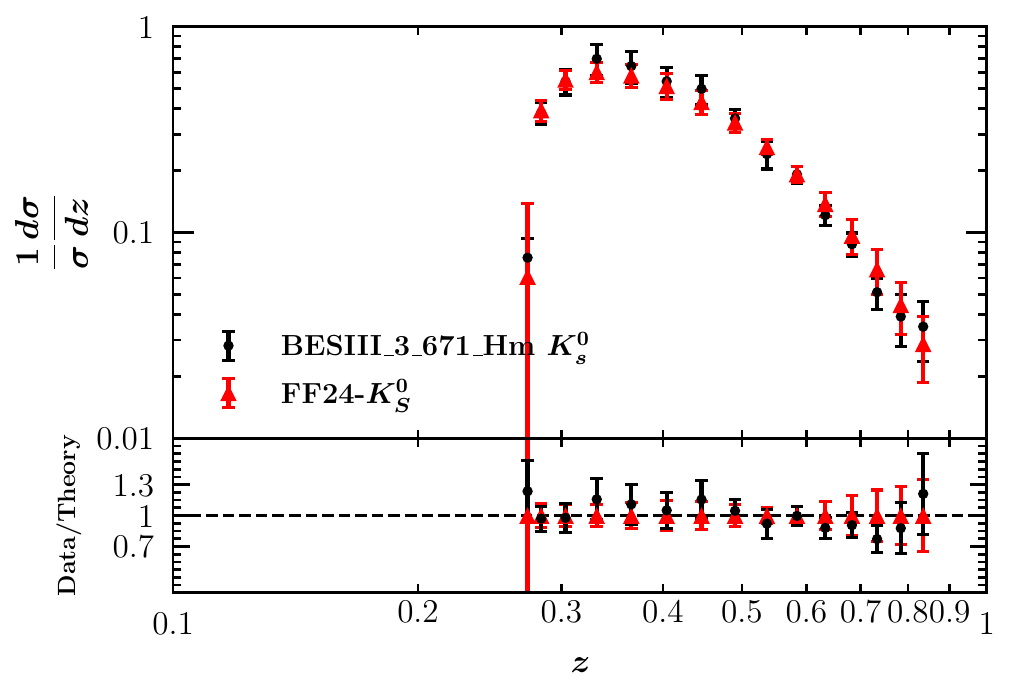}}\\ 	
	\subfloat{\includegraphics[width=0.33\textwidth]{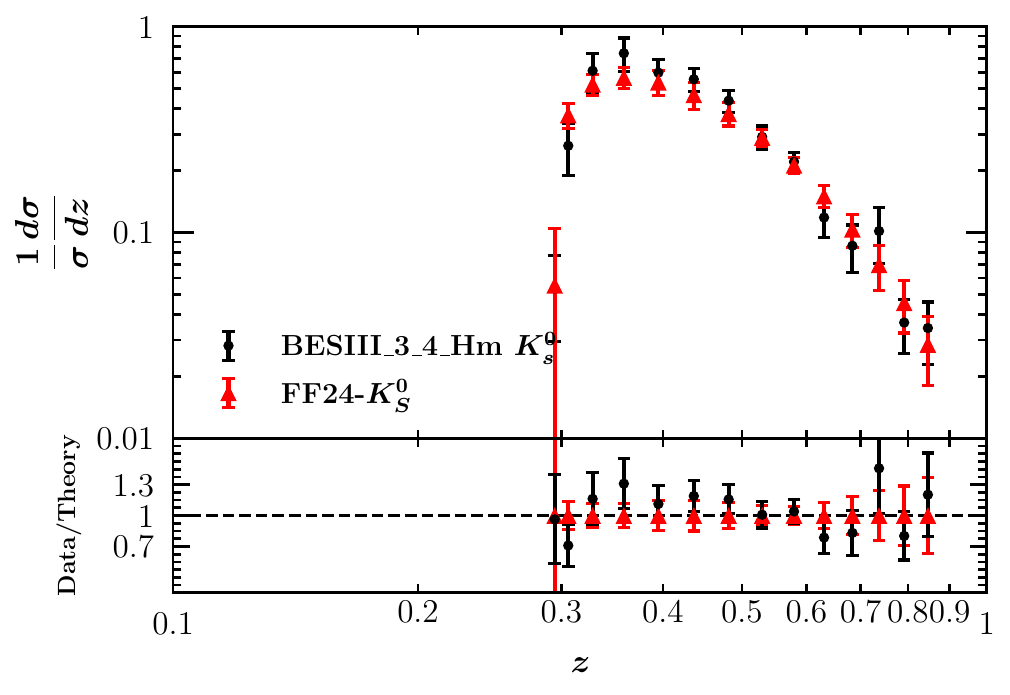}}
	\subfloat{\includegraphics[width=0.33\textwidth]{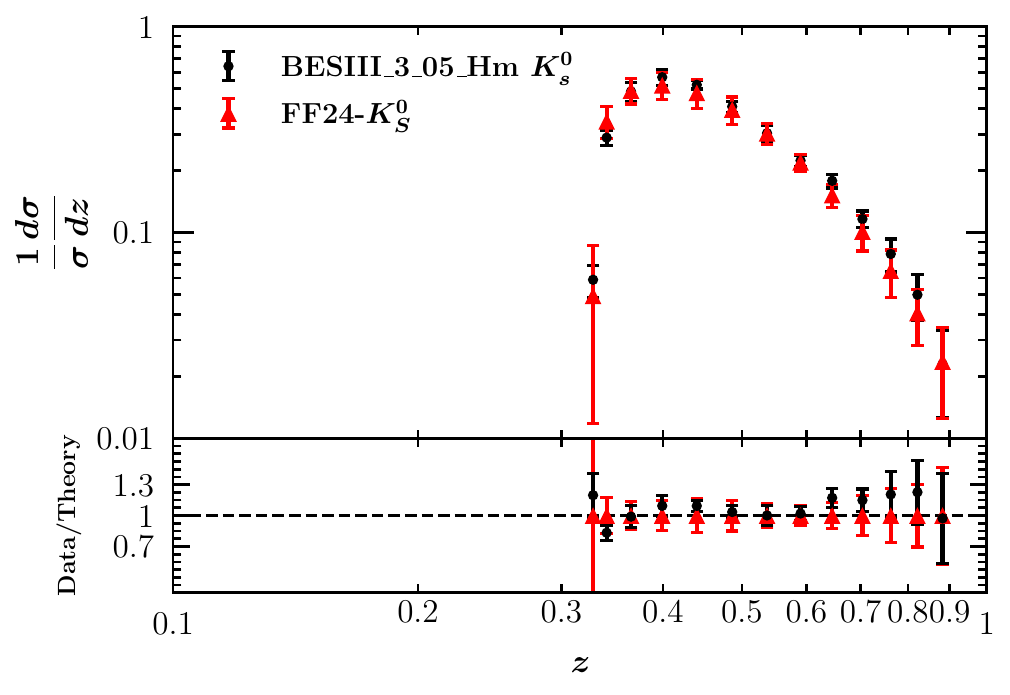}}	
	\subfloat{\includegraphics[width=0.33\textwidth]{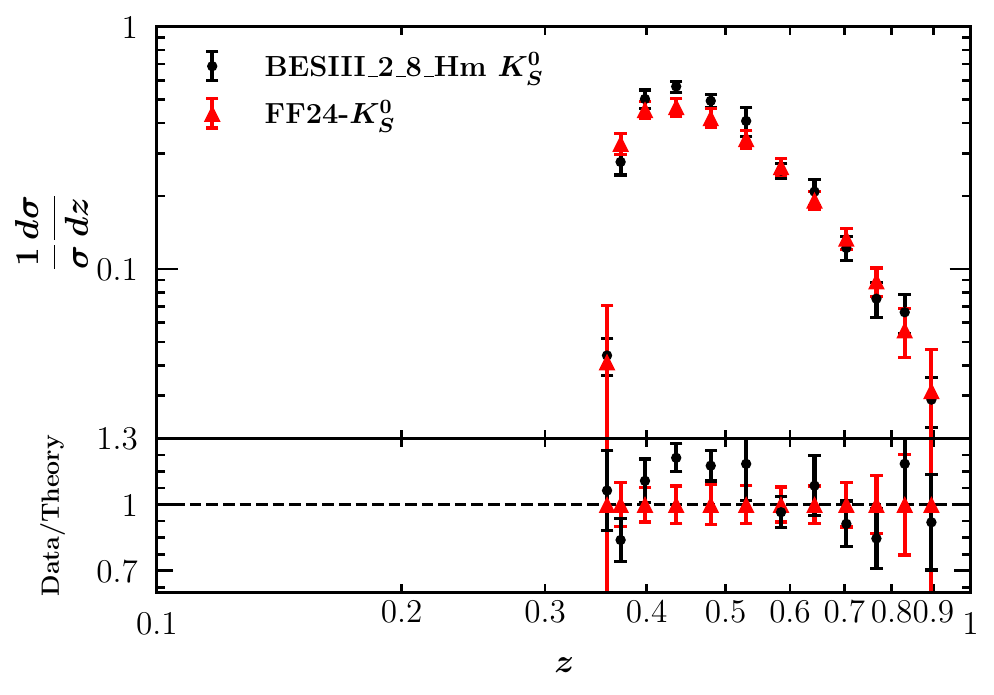}}\\
	\subfloat{\includegraphics[width=0.33\textwidth]{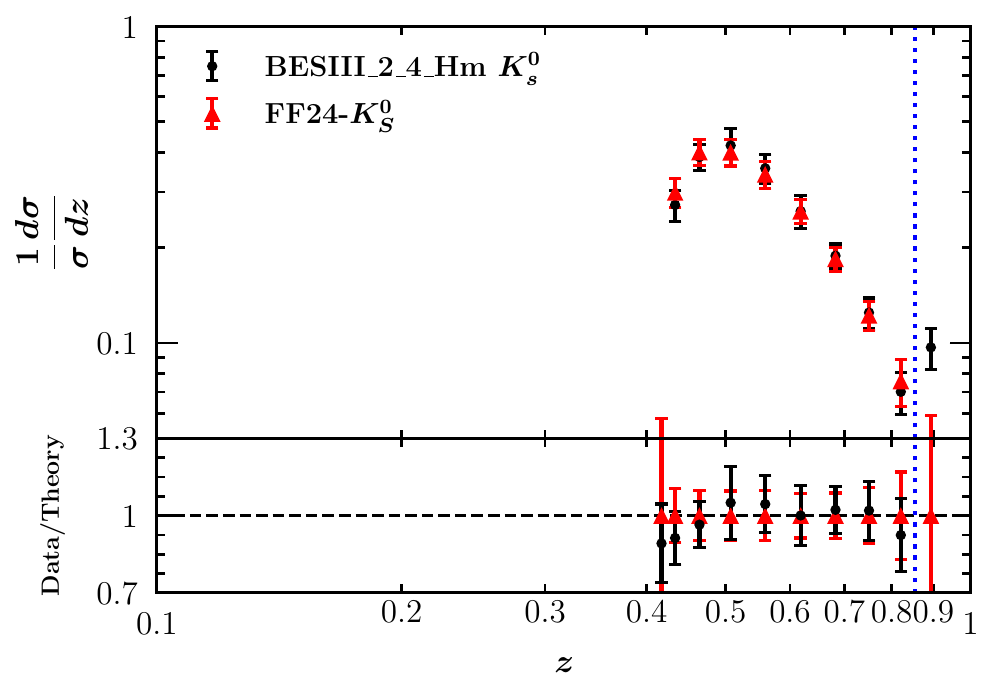}}	
	\subfloat{\includegraphics[width=0.33\textwidth]{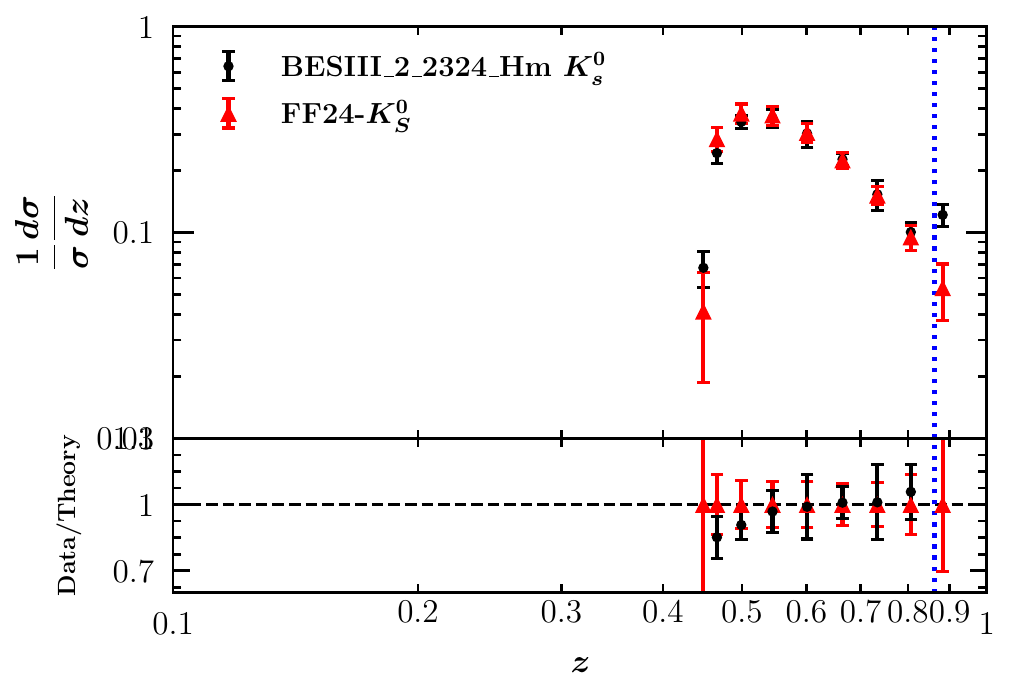}}		
\begin{center}
\caption{ \small 
Same as Fig.~\ref{fig:data1} but for some other selected experimental 
data including \texttt{TASSO14.8}, \texttt{TASSO14}, 
\texttt{BESIII3.671}, \texttt{BESIII3.4}, \texttt{BESIII3.05}, 
\texttt{BESIII2.8}, \texttt{BESIII2.4} and \texttt{BESIII2.2324}. 
}
\label{fig:data3}
\end{center}
\end{figure*}

Some of the data points have been excluded from the fits, as 
explained earlier. In the figures, these points are separated by 
a blue dotted line. Specifically, we have omitted two data points 
from \texttt{ALEPH} and four data points from \texttt{DELPHI} in the 
small $z$ region due to their tension with the rest of the data. 
Additionally, we observed outliers in the last data points of 
\texttt{BESIII2.2324} and \texttt{BESIII2.4}, which deviate from the 
overall trend. Consequently, we decided to exclude them from our 
analysis as well. All data points from the remaining datasets are  
included in our analysis. The $\chi^2$ values listed in 
Table~\ref{tab:datasetsK0s} demonstrate that our predictions are in 
good agreement with the data. A detailed comparison of the $K^0_S$ 
production cross section calculated at NNLO accuracy and the newly 
added \texttt{BESIII} data are presented in Fig.~\ref{fig:data3}. 
As can clearly be seen, very nice agreement is achieved, which is 
consistent with the $\chi^2$ values presented in 
Table~\ref{tab:datasetsK0s}. The incorporation of \texttt{BESIII} data 
contributes significantly to the precision and reliability of our 
new NNLO FF set.

It is worth noting that when datasets are consistent with each other, 
one would typically expect the predictions to have uncertainties of 
a similar magnitude as those of the experimental data. However, if 
there are inconsistencies in the data, uncertainties may decrease to 
account for this discrepancy. This phenomenon, where the combination 
of multiple datasets leads to a reduction in the overall uncertainty, 
is evident in Figs.~\ref{fig:data1}, \ref{fig:data2}, and 
\ref{fig:data3}. Similar observations can be made in Figure 2 of 
\cite{Khalek:2021gxf} and in our previous analysis using MontBlanc 
\cite{Soleymaninia:2022alt}. 

Now we present our $K^0_S$ FFs, denoted as \texttt{FF24-}$K^0_S$, 
obtained from our baseline NLO and NNLO QCD fits and compare them 
with some other FF parametrizations available in the literature. 
Such comparisons are shown in Figs.~\ref{fig:FFsNLO} and 
\ref{fig:FFsN2LO}. For the NLO analysis, we compare with 
\texttt{SAK20}~\cite{Soleymaninia:2020ahn}, 
\texttt{AKK08}~\cite{Albino:2008fy}, and 
\texttt{DSS17}~\cite{deFlorian:2017lwf}. Results from 
\texttt{SAK20}~\cite{Soleymaninia:2020ahn} and 
\texttt{NNFF1.0}~\cite{Bertone:2017tyb} are compared with our 
NNLO fit results.

\begin{figure*}[htb]
	\vspace{0.50cm}
	\centering
	\subfloat{\includegraphics[width=0.33\textwidth]{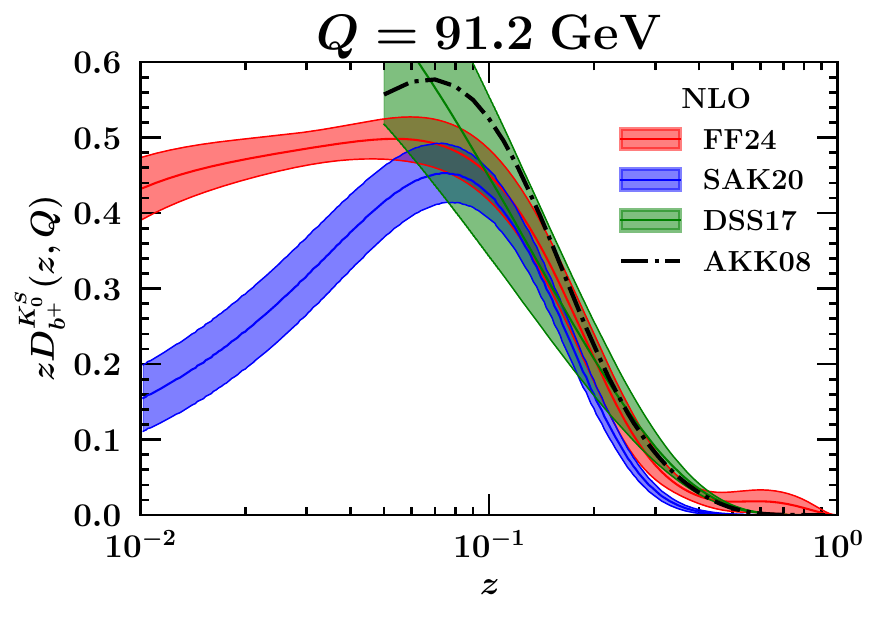}} 	
	\subfloat{\includegraphics[width=0.33\textwidth]{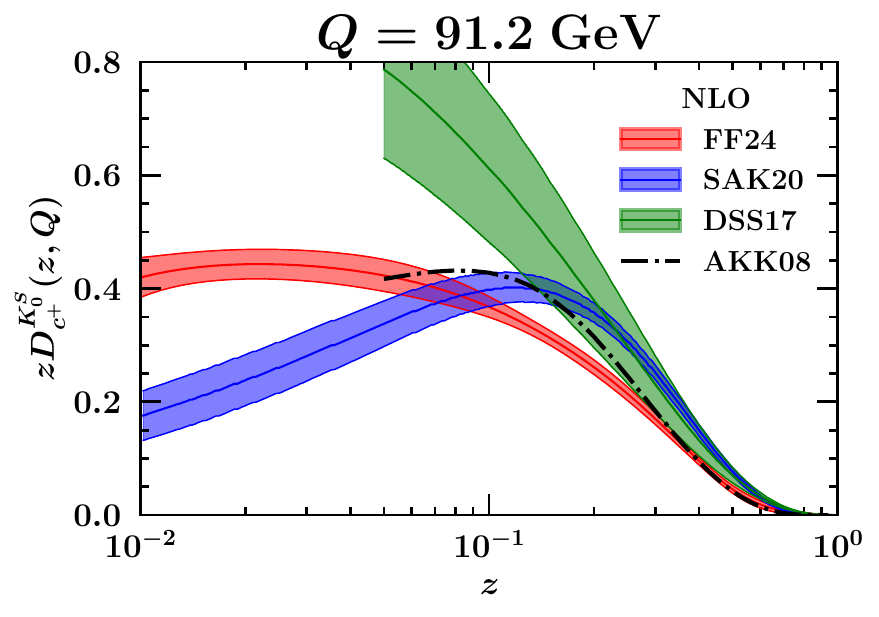}}
	\subfloat{\includegraphics[width=0.33\textwidth]{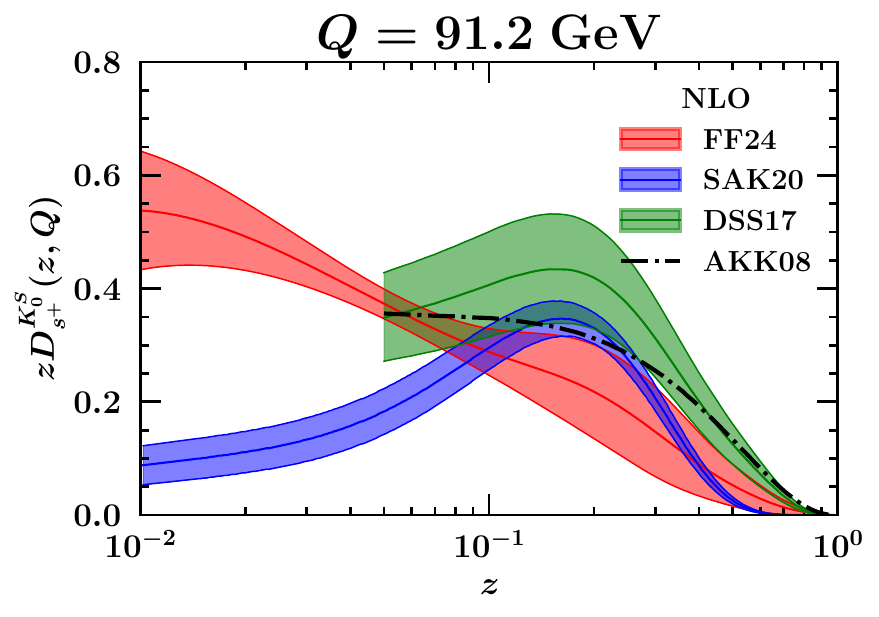}}\\	
	\subfloat{\includegraphics[width=0.33\textwidth]{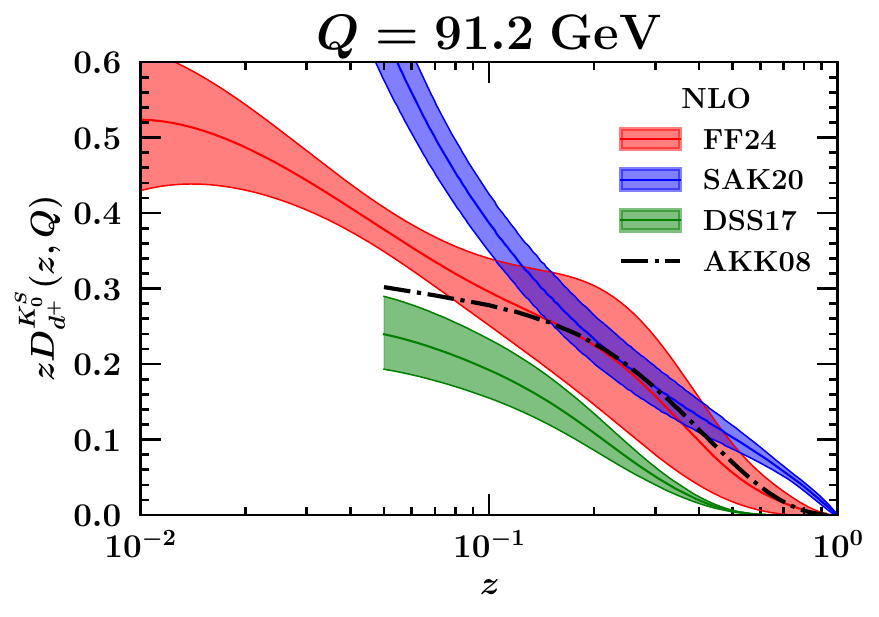}} 
	\subfloat{\includegraphics[width=0.33\textwidth]{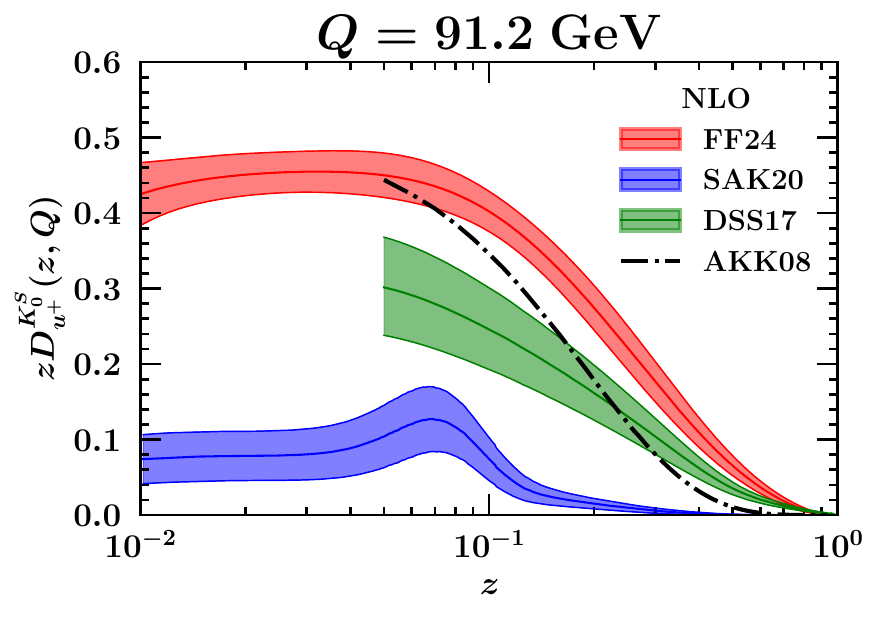}}	
	\subfloat{\includegraphics[width=0.33\textwidth]{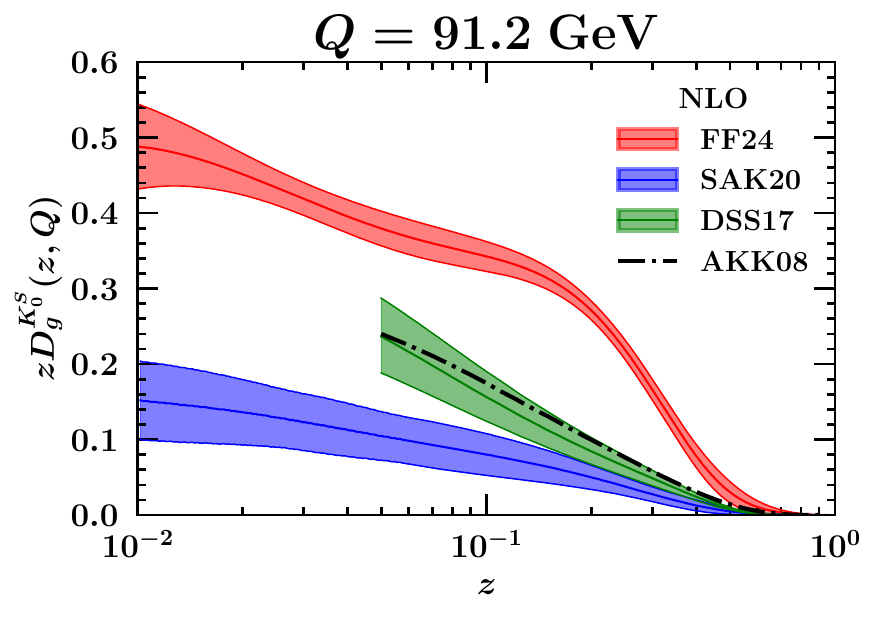}}	
	\begin{center}
\caption{ \small 
The \texttt{FF24-}$K^0_S$ at NLO accuracy obtained for various partons 
at $\sqrt{s}=91.2$~GeV. The shaded bands represent $1\sigma$ uncertainty 
estimates derived from the Monte Carlo method. Results from 
\texttt{SAK20}~\cite{Soleymaninia:2020ahn}, 
\texttt{DSS17}~\cite{deFlorian:2017lwf}, 
and \texttt{AKK08}~\cite{Albino:2008fy} at NLO accuracy, along with 
their $1\sigma$ error bands, are also included for comparison.}
\label{fig:FFsNLO}
\end{center}
\end{figure*}

All four FF sets depicted in Fig.~\ref{fig:FFsNLO} include 
a comparable array of SIA datasets. However, our new {\tt FF24-}$K^0_S$ 
set uniquely integrates the \texttt{BESIII} measurements for the first 
time. Additionally, the {\tt AKK08} FF encompasses $K^0_S$ hadron 
production measurements from $pp(\bar{p})$ collisions.
With the exception of \texttt{DSS17}, hadron mass corrections 
are accounted for in the other three FF sets. Both the {\tt FF24-}$K^0_S$ 
and {\tt SAK20} FFs are computed up to NNLO accuracy.
In contrast, the {\tt AKK08} FFs are determined only up 
to NLO accuracy and available only without an uncertainty estimate.
\texttt{DSS17}~\cite{deFlorian:2017lwf} also presented their 
FFs at NLO accuracy along with uncertainty bands extracted from SIA, 
lepton-nucleon DIS, and proton-proton collision data.
The minimum value for $z$ is considered to be $0.05$ in both 
the \texttt{SAK20} and \texttt{AKK08} analyses. However, 
in the \texttt{FF24-}$K^0_S$ analysis, we incorporated data points 
also for smaller $z$, down to $z \geq 0.013$. This difference 
could result in some variations among the four FF analyses for all 
flavors, in particular in the small-$z$ region, $z<0.05$.

Considering the comparisons presented in Fig.~\ref{fig:FFsNLO}, 
several observations are noteworthy. It is evident that the central 
values of these analyses exhibit different behaviors, particularly 
for the gluon and $u^+$ FFs, even within regions defined by the same 
kinematical cuts. Despite these differences, some similarities emerge, 
such as at large $z$ values for the $b^+$, $c^+$, $s^+$, and $d^+$ FFs. 
In the small $z$ region, where more data points are included 
in the \texttt{FF24-}$K^0_S$ analysis, the FFs are generally 
larger than those of the \texttt{SAK20} FF, except for the gluon  FF. 

Regarding uncertainty bands, there are some comments need to be made 
as well. As shown in Fig.~\ref{fig:FFsNLO}, the uncertainty bands 
for {\tt FF24-}$K^0_S$ FFs are broader compared to those of the 
{\tt SAK20} analysis across nearly all the $z$ region. This difference 
primarily arises from our utilization of the Monte Carlo method to 
accommodate error bands, whereas the {\tt SAK20} analysis employs the 
standard Hessian method. Notably, the uncertainty bands for the 
new {\tt FF24-}$K^0_S$ FFs are wide over regions at small values of 
$z$ due to the insufficient coverage of SIA data points in those areas. 
These wide uncertainty bands are particularly pronounced for 
the $u^+$ and $s^+$ FFs. Upon comparison with \texttt{DSS17}, 
it becomes evident that their analysis yields larger error bands 
across nearly all FFs, except for the $d^+$ FFs.

We now shift our focus to the comparison of the \texttt{FF24-}$K^0_S$ 
at NNLO accuracy with those of \texttt{SAK20}~\cite{Soleymaninia:2020ahn} 
and \texttt{NNFF1.0}~\cite{Bertone:2017tyb}. These comparisons are 
illustrated in Fig.~\ref{fig:FFsN2LO}. Considering the central values 
of these analyses, one can observe different behaviors for all parton 
species, even within regions defined by the same kinematical cuts.
For the case of $b^+$, $c^+$, and $s^+$ FFs, the central values 
of \texttt{FF24-}$K^0_S$ are smaller than those of 
\texttt{SAK20}~\cite{Soleymaninia:2020ahn} and 
\texttt{NNFF1.0}~\cite{Bertone:2017tyb}. 
However, for the $u^+$ and gluon FFs, although they are larger in 
comparison to \texttt{SAK20}, they are still smaller than those of 
\texttt{NNFF1.0}. In terms of uncertainty bands, it is noticeable 
that the error bands of \texttt{FF24-}$K^0_S$ are smaller than those 
of \texttt{NNFF1.0} for almost all parton species and across all 
regions of $z$, particularly for the small-$z$ region $z<0.1$. 
When comparing with \texttt{SAK20}, it can be concluded that smaller 
error bands are achieved, except for the case of $d^+$ and $s^+$ FFs.

\begin{figure*}[htb]
	\vspace{0.50cm}
	\centering
	\subfloat{\includegraphics[width=0.33\textwidth]{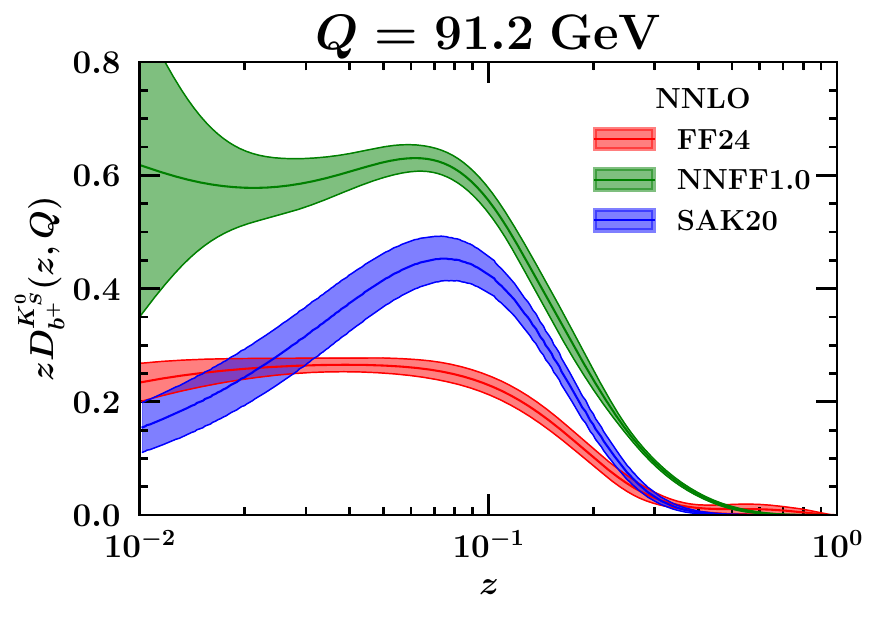}} 	
	\subfloat{\includegraphics[width=0.33\textwidth]{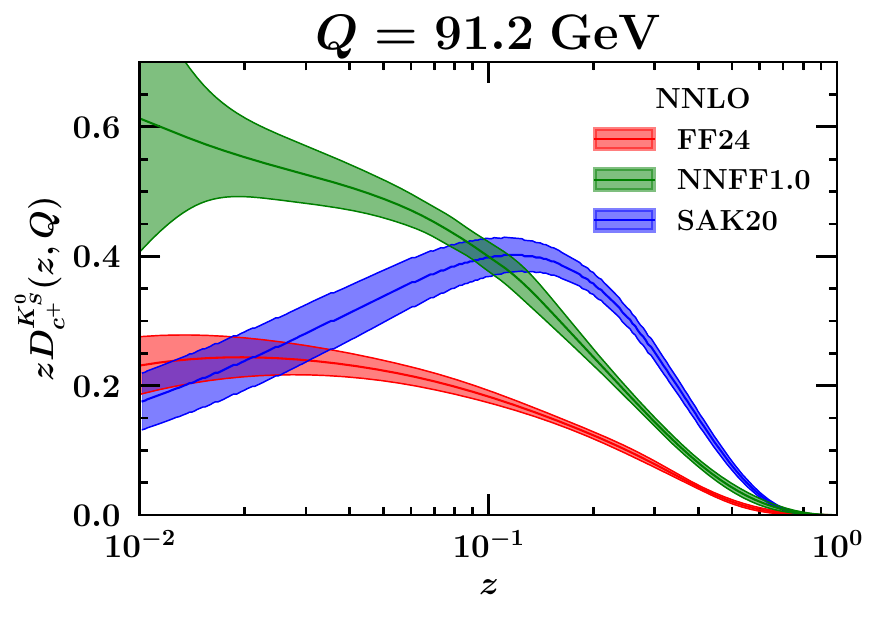}}
	\subfloat{\includegraphics[width=0.33\textwidth]{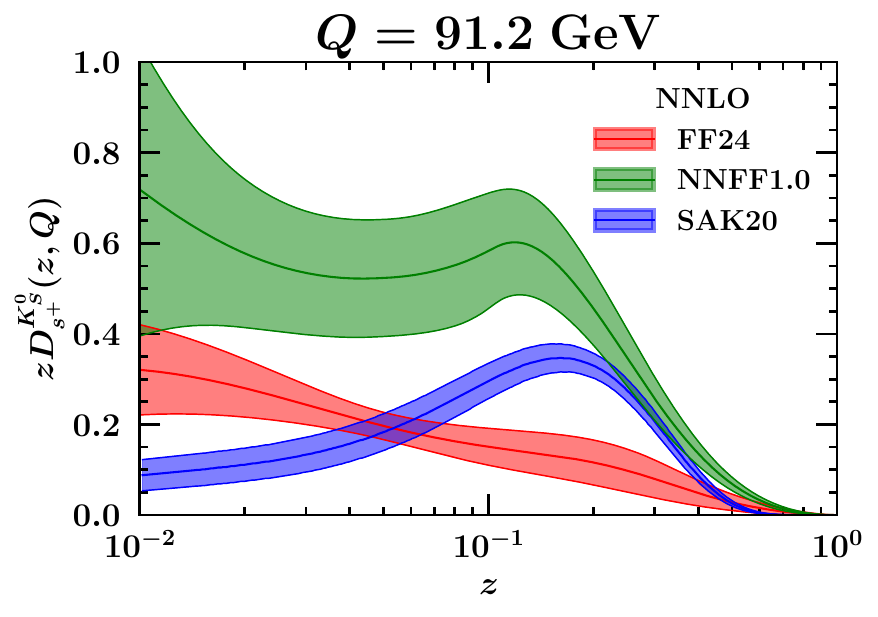}}\\	
	\subfloat{\includegraphics[width=0.33\textwidth]{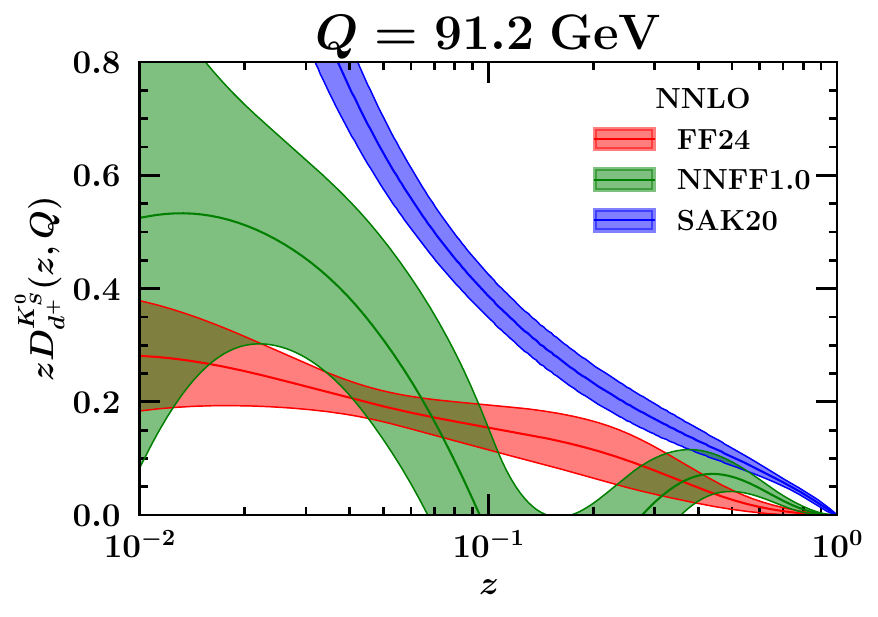}} 
	\subfloat{\includegraphics[width=0.33\textwidth]{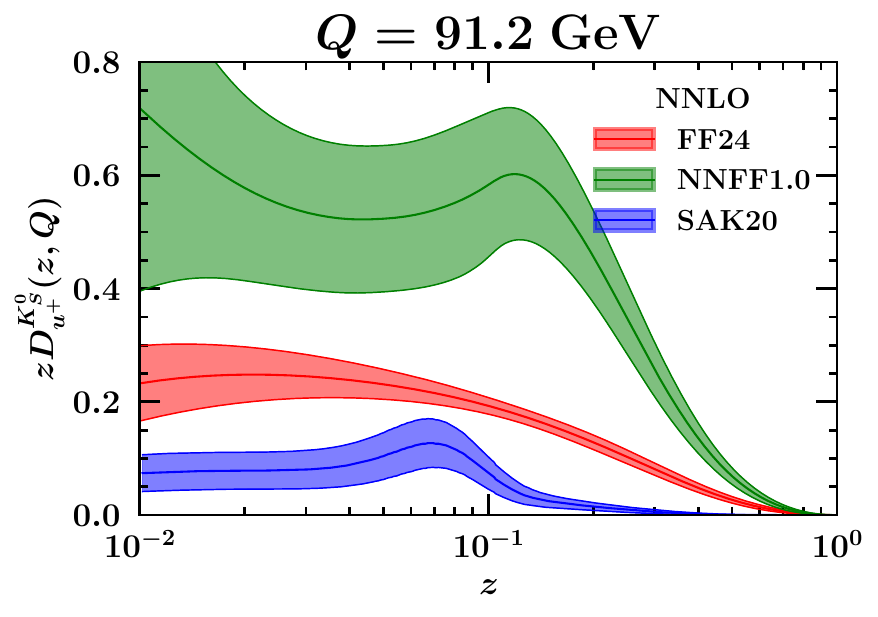}}	
	\subfloat{\includegraphics[width=0.33\textwidth]{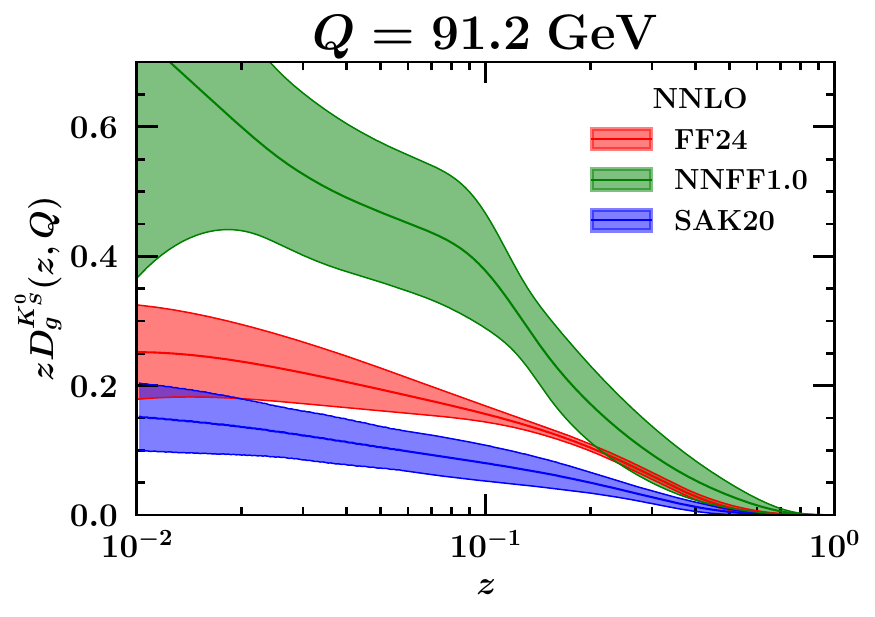}}	
	\begin{center}
\caption{ \small 
Same as Fig.~\ref{fig:FFsNLO}, but this time for the 
\texttt{FF24-}$K^0_S$ at NNLO accuracy. 
Results from \texttt{SAK20}~\cite{Soleymaninia:2020ahn} and 
\texttt{NNFF1.0}~\cite{Bertone:2017tyb}, along with their error bands 
at NNLO accuracy, are also included for comparison. }
\label{fig:FFsN2LO}
\end{center}
\end{figure*}

In order to see how the \texttt{BESIII} data for $K^0_S$ 
production affect the determination of FFs, we compare in 
Fig.~\ref{fig:SAK20} data with predictions obtained from 
both the previous \texttt{SAK20} FFs and the new 
\texttt{FF24-}$K^0_S$ FFs. We observe large differences 
between data and \texttt{SAK20} predictions over the full  
range of $z$ and for all energies. The theory predictions based 
on \texttt{SAK20} FFs are not in good agreement with the 
\texttt{BESIII} data. In particular, the increase of cross 
section predictions from \texttt{SAK20} towards low values 
of $z$ is not seen in the data. We believe that the main 
reason for this difference is that the \texttt{SAK20} FFs 
had been extracted from high-energy SIA data and the 
predictions using QCD-backward evolution of FFs from high 
to low energies do not match the experimental results. We 
conclude that including \texttt{BESIII} data improve the fit 
for FFs in particular at low $z$.

\begin{figure*}[htb]
	\vspace{0.50cm}
	\centering
	\subfloat{\includegraphics[width=0.33\textwidth]{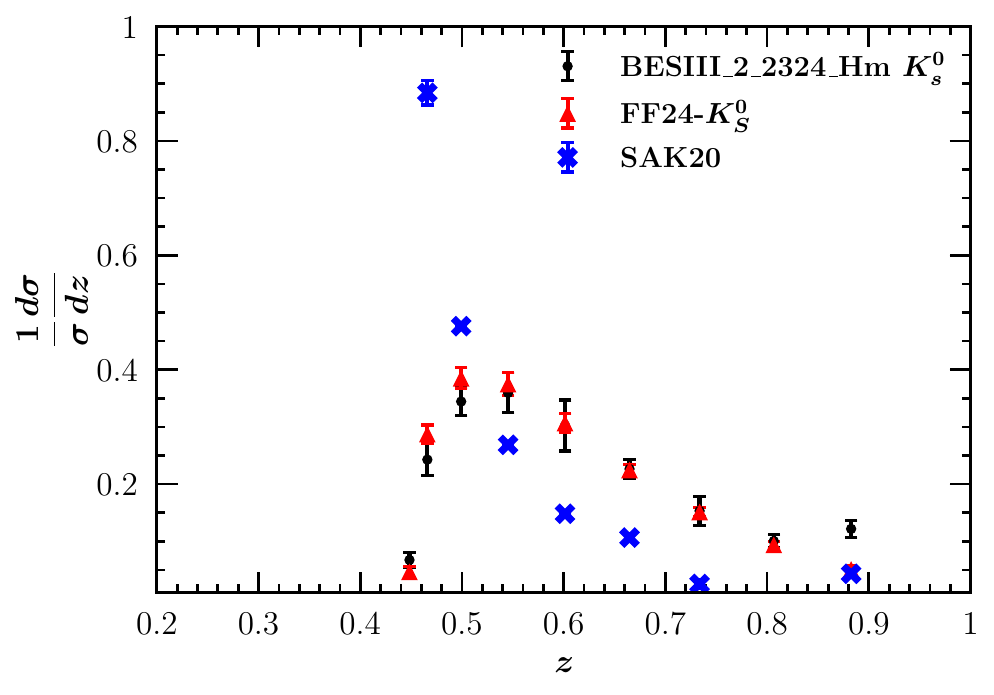}}	
	\subfloat{\includegraphics[width=0.33\textwidth]{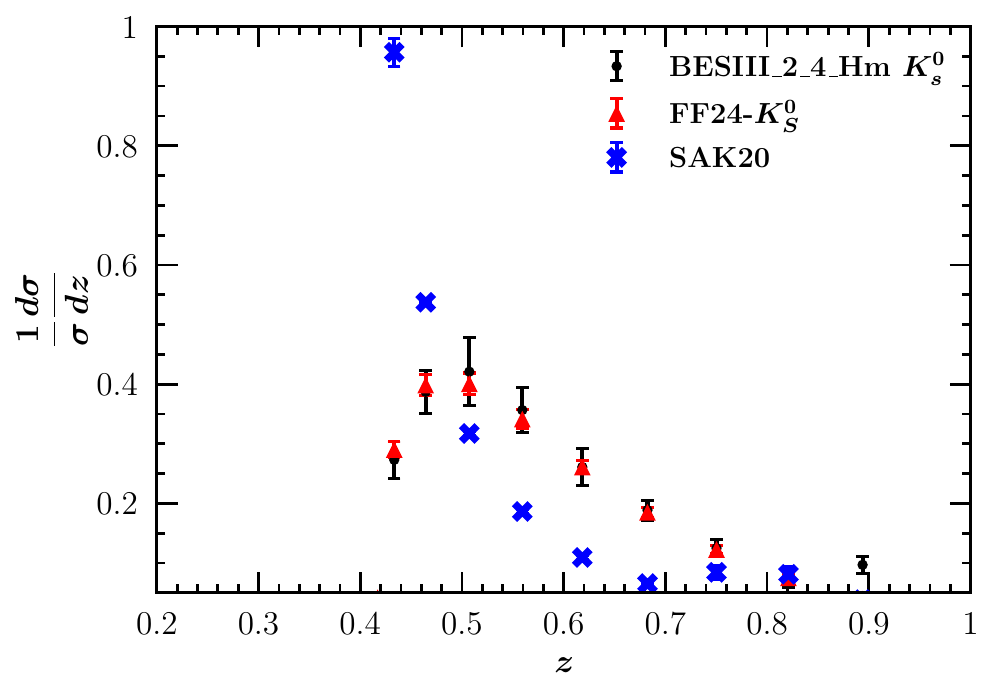}}\\
	\subfloat{\includegraphics[width=0.33\textwidth]{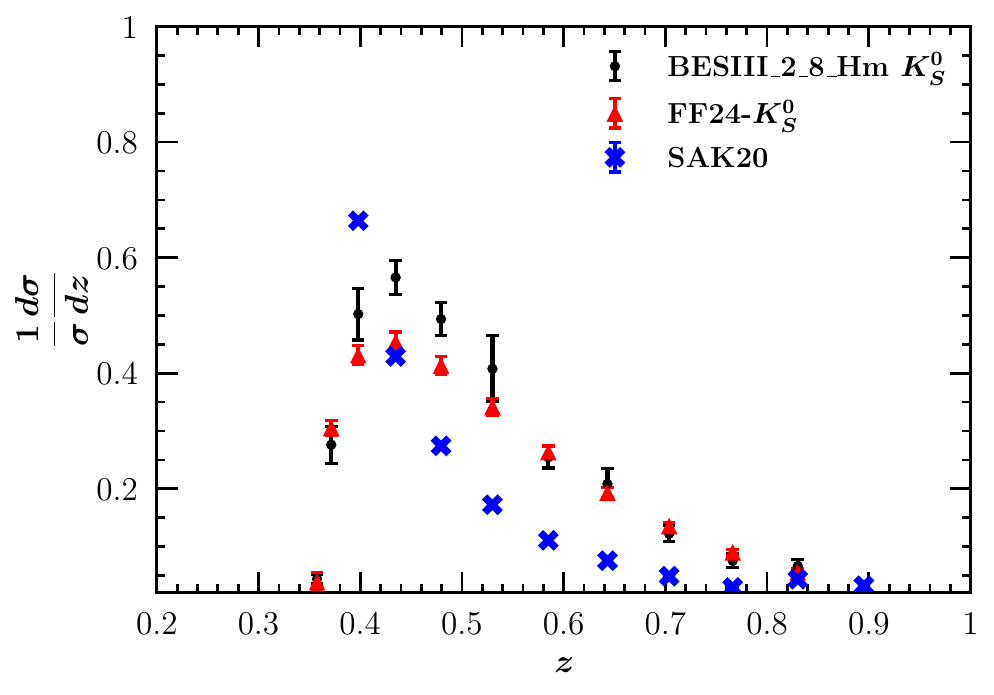}}	
	\subfloat{\includegraphics[width=0.33\textwidth]{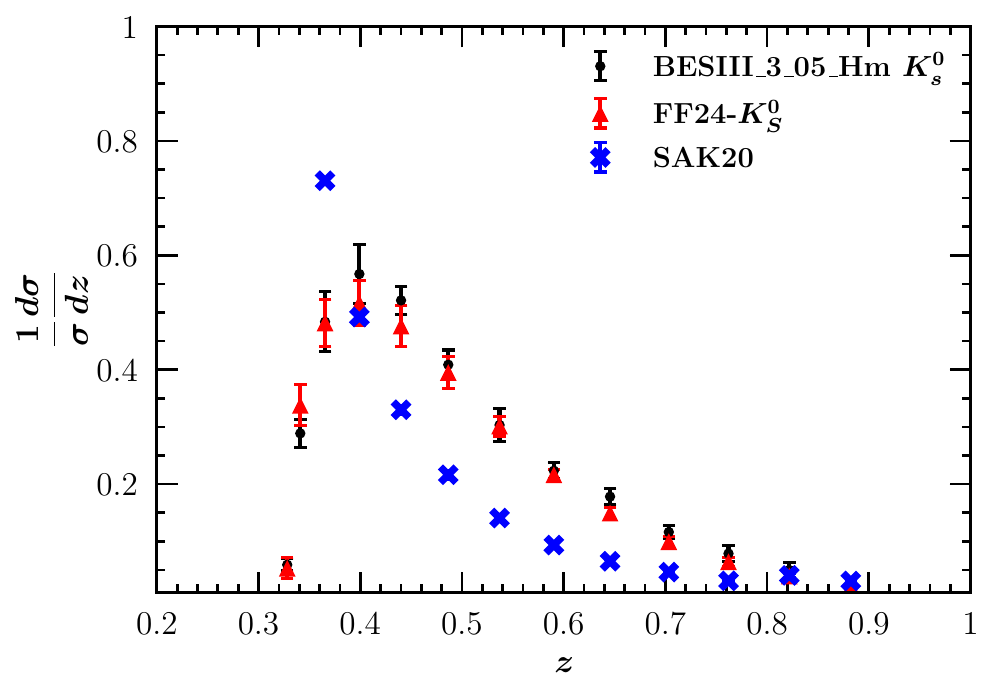}}\\
	\subfloat{\includegraphics[width=0.33\textwidth]{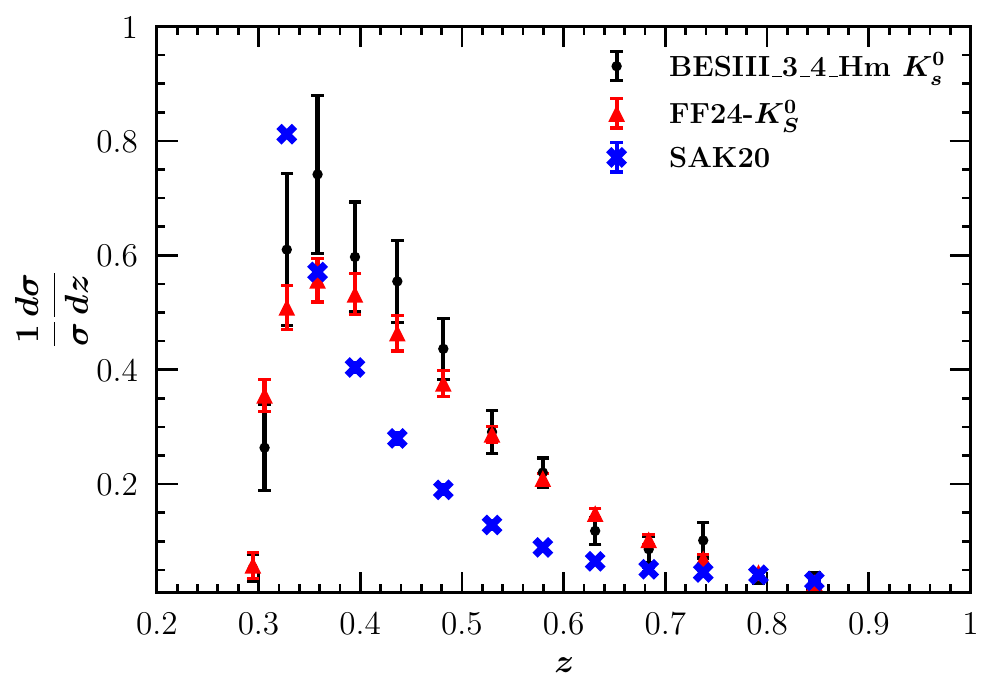}}	
	\subfloat{\includegraphics[width=0.33\textwidth]{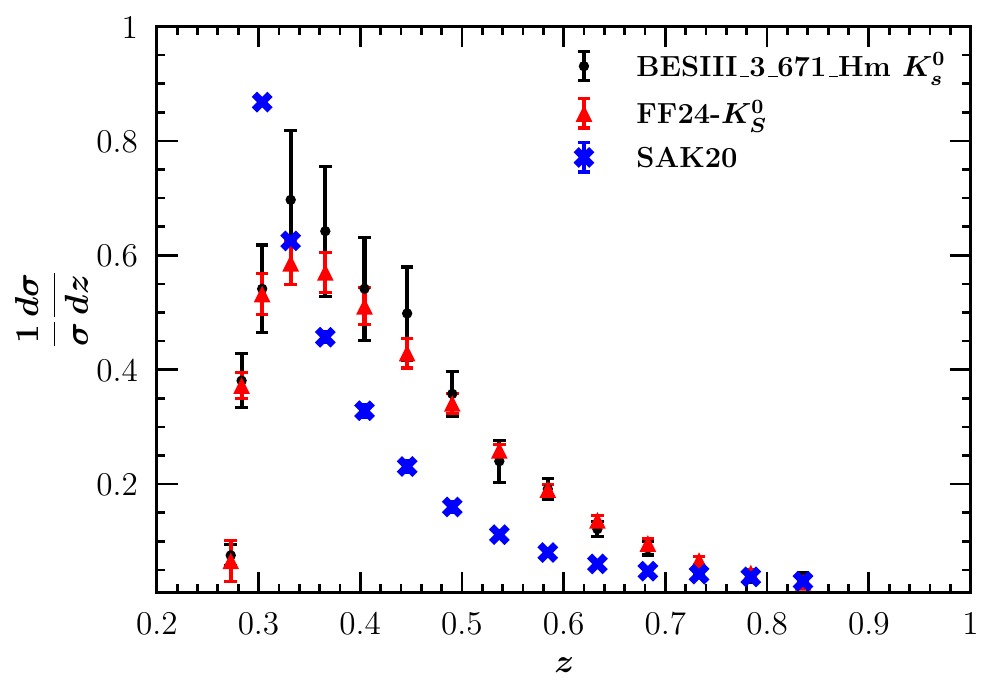}}		
	\begin{center}
		\caption{ \small 
		Comparison between the predictions for cross sections 
		of $K^0_S$ production from 
		\texttt{SAK20}~\cite{Soleymaninia:2020ahn} and the 
		present analysis at NLO accuracy and the measurements 
		from the \texttt{BESIII} experiment. 
		}
		\label{fig:SAK20}
	\end{center}
\end{figure*}

In summary, our approach highlights two key improvements: First, 
the integration of the latest experimental data from {\tt BESIII}, 
which enriches the dataset and enhances the reliability of our results. 
Second, the adoption of neural networks in the fitting procedure 
introduces a novel technique to refine the extraction of FFs. 
This innovative combination of advanced methodologies contributes 
to a more accurate determination of $K^0_S$ FFs, offering 
improved precision and reliability for the prediction of measurements 
at future high-energy scattering experiments. 

%
\section{Summary and outlook}
\label{Conclusion}
%

In summary, the main goal of this paper is to introduce new sets of 
fragmentation functions for $K^0_S$, entitled {\tt FF24-}$K^0_S$, 
obtained from a QCD analysis including NNLO corrections. Our analysis 
relies on extensive experimental data, incorporating measurements of 
$K^0_S$ production in the SIA process and including the most recent 
experimental data reported by the \texttt{BESIII} Collaboration.

The \texttt{BESIII} measurement covers collision energies spanning 
from $\sqrt{s}=2.2324$ to $\sqrt{s}=3.6710$~GeV below the range used 
in earlier FF fits. The findings presented in this study hold particular 
significance as they contribute valuable insight, especially in this 
new energy range below $\sqrt{s} < 10$~GeV where precise $e^+e^-$ 
annihilation data have been scarce before. By analyzing normalized 
differential cross sections across this energy range, we have 
successfully addressed critical gaps in our understanding of 
$K^0_S$ hadron production.

Incorporating hadron mass corrections particularly in datasets with 
low values of energy and $z$, is crucial. In our analyses, we have 
applied $K^0_S$ mass corrections which allowed us to incorporate 
low-$z$ data points, excluding only 8 data points with $z_{\text{min}} 
= 0.013$. While small-$z$ resummation is not explicitly addressed in 
our present analysis, we acknowledge its importance and plan to explore 
it in future investigations.

We have utilized neural networks to parameterize the fragmentation 
functions, aiming to minimize theoretical bias. The NN parameters 
are tuned to fit the SIA data. Our approach used the recent publicly 
available package {\tt MontBlanc} for FF parametrization, evolution, 
and SIA production cross-section calculation. Moreover, we employ a 
Monte Carlo sampling method to propagate experimental uncertainties 
into the fitted FFs and determine uncertainties for both FFs and 
associated observables.

Both the overall $\chi^2$ and the $\chi^2$ values for individual 
datasets used in the fitting procedure are satisfactory. This is 
evident in the excellent agreement observed between the experimental 
data and the corresponding theoretical predictions. However, upon 
comparison with two other determinations, namely {\tt SAK20} and 
{\tt AKK08}, some notable differences were identified in the resulting 
{\tt FF24-}$K^0_S$ set. In this work, we focus exclusively on SIA 
observables to accurately determine the quark FFs of $K^0_S$. However, 
these measurements provide limited information about the gluon-to-hadron 
FFs since they are only accessible starting at NLO. Consequently, 
significant differences arise between our findings and those of other 
results available in the literature, such as {\tt AKK08}, where 
$K^0_S$ production was investigated using also collider data. To gain 
a better understanding of the gluonic contribution, we plan to extend 
our analysis to include data from the production of $K^0_S$ particles 
in proton collisions. Previous studies suggest that this will be 
particularly important for a better determination of the role of 
gluons in the overall picture.

The resulting \texttt{FF24-}$K^0_S$ sets at NLO and NNLO accuracy, 
along with their associated uncertainties, are made available in 
the standard {\tt LHAPDF} format for broader utilization in the 
scientific community\cite{ff24}.

%
\begin{acknowledgments}
%

The authors gratefully acknowledge many helpful discussions and comments 
by Valerio Bertone that elevated and enriched the paper significantly.
Hamzeh Khanpour, Maryam Soleymaninia and Hadi Hashamipour thank the 
School of Particles and Accelerators, Institute 
for Research in Fundamental Sciences (IPM) for financial support of 
this project.  
Hamzeh Khanpour also appreciates the financial support from NAWA
under grant number BPN/ULM/2023/1/00160 and from the IDUB programme
at the AGH University. 
This work was also supported in part
by the Deutsche Forschungsgemeinschaft (DFG, German Research
Foundation) through the funds provided to the Sino-German Collaborative
Research Center TRR110 ``Symmetries and the Emergence of Structure 
in QCD'' (DFG Project-ID 196253076 - TRR 110).  The work of 
UGM was also supported in part by the Chinese
Academy of Sciences (CAS) President's International
Fellowship Initiative (PIFI) (Grant No.\ 2018DM0034).

\end{acknowledgments}

\clearpage


%

\end{document}